\documentclass[aps,prd,twocolumn,nofootinbib,amsmath,amssymb,showpacs,longbibliography]{revtex4-1}

\usepackage{graphics}
\usepackage{color}
\usepackage{hyperref}

\usepackage{url}
\usepackage{breakurl}

\usepackage[utf8]{inputenc}

\usepackage{lipsum}
\begin{document}
%%%%%%%%%%%%%%%%%%%%%%%%%%%%% BIBITEX %%%%%%%%%%%%%%%%%%%%%%%%%%%%%%%%%%%%
% \bibliographystyle{apsrev4-1}%unsrt}
 
\title{Multiplicity fluctuations near the QCD critical point}

\author{M. {\sc Hippert}}
\email{hippert@if.ufrj.br}
\author{E. S. {\sc Fraga}}
\email{fraga@if.ufrj.br}
\affiliation{Instituto de F\'\i sica, Universidade Federal do Rio de Janeiro,
Caixa Postal 68528, 21941-972, Rio de Janeiro, Rio de Janeiro, Brazil}

\date{\today}

%%%%%%%%%%%%%%%%%%%%%%%%%%%%%%%%%%%%%%%%%%%%%%%%%%%%%%%%%%%%%%%%%%%%%%%%%%%%%%

% \pacs{25.75.Nq, 11.10.Wx, 12.39.Fe, 64.60.Q-}

%%%%%%%%%%%%%%%%%%%%%%%%%%%%%%%%%%%%%%%%%%%%%%%%%%%%%%%%%%%%%%%%%%%%%%%%%%%%%%

\begin{abstract}
Statistical moments of particle multiplicities in heavy-ion collision experiments are an important probe in the exploration of the phase diagram of 
strongly interacting matter and, particularly, 
in the search for the QCD critical end point. In order to appropriately interpret experimental measures of these moments, however, it is necessary to 
understand the role of experimental limitations, as well as background contributions,  
providing expectations on how critical behavior should be affected by them. 
We here present a framework for calculating moments of particle multiplicities in the presence of correlations of both critical and spurious origins.  
We also include effects from resonance decay and a limited acceptance window, as well as detector efficiency. 
Although we focus on second-order moments, for simplicity, an extension to higher-order moments is straightforward.
\end{abstract}

\maketitle

%%%%%%%%%%%%%%%%%%%%%%%%%%%%%%%%%%%%%%%%%%%%%%%%%%%%%%%%%%%%%%%%%%%%%%%%%%%%%%

\section{Introduction}

The QCD phase diagram in the temperature {versus} baryonic chemical potential plane is marked by very distinct regimes, characterized by different (approximate) symmetries and 
degrees of freedom. However, the nature of these regimes and the transitions between them is still not settled.  
While first-principle calculations face the breakdown of perturbation theory at low and intermediate energy scales and  
lattice calculations remain limited to relatively low baryon  densities, where an analytic crossover is found~\cite{Aoki:2006we, Borsanyi:2010cj}, 
effective models suggest a first-order phase transition at high densities. 
The first-order coexistence line, related to the restoration of chiral symmetry, should end at lower densities in a  second-order critical end point of unknown location~\cite{Stephanov:2007fk,Stephanov:2004wx,Rajagopal:1995bc}.

The possibility of experimentally probing this point with current technology and the very peculiar singular 
behavior associated with a second-order  phase transition raise very appealing hopes of accessing reliable 
information on the main features of the phase diagram of strongly interacting matter, 
which have motivated considerable experimental and theoretical efforts. 
The search for the QCD critical point is currently ongoing at the RHIC Beam Energy Scan program~\cite{Luo:2015doi,Luo:2015ewa,Xu:2016mqs}.

On the other hand, notwithstanding its very characteristic features in an ideal scenario, the observation or exclusion of the 
critical point in the noisy and dynamic context of  heavy-ion collision experiments remains an extremely challenging task. 
Whereas the main probe in the pursuit for such a point is the increase of long-range fluctuations caused by a diverging 
 correlation length in its neighborhood~\cite{Stephanov:1999zu,Stephanov:2008qz,Athanasiou:2010kw}, the short lifetime and small size of the formed system will 
considerably limit its growth~\cite{Berdnikov:1999ph,Braun:2005fj,Kiriyama:2006uh,Palhares:2009tf,Fraga:2011hi,Mukherjee2015,Mukherjee2016,Hippert2016}. 
Moreover, collisions are not perfectly controlled and are affected by initial-state fluctuations, so that spurious 
sources of fluctuations could possibly hide or mimic the expected signatures~\cite{Stephanov:1999zu,Hippert2016,Gorenstein2015,Sahoo2013,Herold2016,Luo2013a}.   
Finally, the medium which is formed in experiments undergoes very fast expansion and cooling, followed by freeze-out and rescattering, 
which may also have a strong influence upon critical behavior~\cite{Herold2016,Antoniou2007,Antoniou2008,Stephanov2010}. 
It is thus clear that care should be taken when comparing naive theoretical expectations with actual measurements.  
Such difficulties  call for the use of observables that can serve as robust probes of criticality. 
The search for reliable signatures  
demands a correct understanding of both critical fluctuations and relevant background contributions, 
as well as their dependence on center-of-mass energy, collision centrality and kinematic cuts~\cite{Sahoo2013,Fraga:2011hi,Bzdak2012,Ling2016}. 

In this paper we introduce a simple, yet reasonably general framework for understanding how both critical and 
background fluctuations affect different fluctuation measures. 
For that purpose, we adopt a simplified model of long-range fluctuations~\cite{Stephanov:2007fk,Stephanov:1999zu,Hippert2016}. 
We also show how kinematic cuts and effects from resonance decay can be included and how they affect our results. 
Since we restrict ourselves to equilibrium and isotropy assumptions, we focus on multiplicity fluctuations. 
This simple framework provides a useful tool for understanding multiplicity fluctuations near the critical point. 

The outline of the paper is as follows. 
In Sec.~\ref{secFluct}, we explain our models for critical and spurious fluctuations and discuss the growth of the correlation length 
near the critical point. In Sec.~\ref{secMethod}, we describe a method for analytically calculating multiplicity fluctuation measures in an 
equilibrated gas of quasi-free particles.  
Sections~\ref{secKC} and \ref{secRes} discuss effects from a limited acceptance window and resonance decay, respectively. 
Finally, Sec.~\ref{secResult} summarizes our results, while Sec.~\ref{secConclusion} presents final remarks. 

\section{Fluctuations near the critical point}
\label{secFluct}
\subsection{Critical fluctuations}
\label{secCritFluct}

Inspired by Refs.~\cite{Stephanov:2007fk,Stephanov:1999zu}, we describe critical fluctuations as fluctuations of an order parameter $\sigma(x)$, with a probability distribution 
\begin{equation}
 \mathcal{P}[\sigma] \sim e^{-\Omega[\sigma]/T}\,,
\label{eqPsigma2}
\end{equation}
and, for small fluctuations, we approximate the effective free energy $\Omega[\sigma]$ as 
\begin{equation}
\Omega[\sigma] \approx \int d^3 x \, \left\{ \dfrac{1}{2} (\nabla \sigma)^2 + \dfrac{1}{2} m_\sigma^2\, \sigma^2 \right\}\,,
 \label{effPot2loc}
\end{equation}
where we have imposed $\langle \sigma \rangle = 0$. 
The mass $m_\sigma$ can be shown to be the inverse of the correlation length $\xi$ and vanishes at the critical point, 
rendering long-wavelength fluctuations essentially costless. 

Thus, as the critical point is approached, long-range fluctuations get increasingly strong until, for very small values 
of $m_\sigma = \xi^{-1}$, they get too large and  Eq.~(\ref{effPot2loc}) loses validity altogether. 
For the sake of simplicity, since long wavelengths dominate, we look exclusively at fluctuations of $\sigma_0 := \int d^3 x \, \sigma(x)/V$. 
Integrating over shorter wavelengths in~(\ref{eqPsigma2}) allows us to define a probability distribution for $\sigma_0$,  
\begin{equation}
 P(\sigma_0) \sim e^{- \frac{V}{T}\frac{m_\sigma^2}{2} \sigma_0^2}\,.
\label{eqPsigma02}
\end{equation}
Non-Gaussian fluctuations, which are important probes in the search for the critical point, can be described by including in~(\ref{effPot2loc}) 
terms of higher order in $\sigma$, 
which would also extend the validity of the model~\cite{Stephanov:2007fk}.  

Equation~(\ref{eqPsigma02}) describes global fluctuations of the order parameter in equilibrium in some region of the phase diagram around the critical end point 
where fluctuations are not too large and a classical treatment for the order parameter is valid. 
It neglects dynamical effects which might make global correlations unattainable in the relevant time scales and provides only a crude model. 
However, this simplification enables us to make both simulations and analytical calculations in a relatively simple fashion, even in the case of
non-Gaussian fluctuations, providing an optimistic, yet valuable tool for understanding how different effects affect critical fluctuations. 
The off-equilibrium evolution of the order parameter is tackled, for instance, in Refs.~\cite{Berdnikov:1999ph,Mukherjee2015,Mukherjee2016,Herold2016}.

A connection with observables requires coupling the order parameter to particle fields. 
We consider linear couplings in $\sigma_0$ which enter the Lagrangian as mass corrections for both protons and pions, 
\begin{equation}
 \mathcal{L}_{int} = G\, \sigma_0\; \vec \pi \cdot \vec \pi + g\, \sigma_0\; \bar\psi_p\, \psi_p\,,
\label{eqLagInt}
\end{equation}
where the value of $G$ can be estimated as $\sim300$~MeV near the critical point by symmetry arguments~\cite{Stephanov:1999zu}, but  
only the vacuum value of $g \sim 10$ is known, from the linear sigma model~\cite{GellMann:1960np}.

In the absence of reliable information on its value as a function 
of collision energy or temperature and chemical potential, we treat $\xi = m_\sigma^{-1}$ as an input to our calculations. 
This correlation length, whose size determines the strength of the fluctuations~\cite{Stephanov:2007fk,Stephanov:1999zu}, is not determined solely by 
an equilibrium equation of state or static universality class exponents, 
but is also influenced by both finite-size effects and dynamics, with the latter being expected to dominate~\cite{Berdnikov:1999ph}. 
We consider limitations to the growth of $\xi$ from critical slowing down as discussed in Ref.~\cite{Berdnikov:1999ph}, 
with the evolution of $\xi(t)$ in the proper-time $t$ given by the ansatz equation
\begin{equation}
 \dfrac{{d} \xi}{{  d} t} = A\; \left(\dfrac{\xi}{\xi_0} \right)^{2-z}\,\left(\dfrac{\xi_0}{\xi} - \dfrac{\xi_0}{\xi_{eq}(t)}\right)\, ,
\label{modBerdnikov1}
\end{equation}
where $\xi_{eq}(t) = \xi_0 \;|{t}/{\tau} |^{-\nu/\beta \delta}$ and 
$\alpha = 0.11$, $\nu=0.63$, $z=2 + \alpha/\nu$, $\beta=0.326$ and $\delta=4.80$, coming from universality class arguments~\cite{ZinnJustin:1998ci, Guida:1996ep, Hohenberg:1977ym}.
The nonuniversal dimensionless parameter $A$ cannot be directly estimated, but is limited by causality, depending on the combination $x= \tau/\xi_0$, 
where $\tau$ defines a cooling time scale and $\xi_0$ is the typical value of the correlation length at which universal behavior sets in~\cite{Hippert2016}. 
However, we believe our previous estimate for the time spent in the critical region,  $\tau = 5.5$ fm, to be overly optimistic, comparable to the lifetime of the system~\cite{Hippert2016}. 
This parameter strongly influences %maximum allowed value of parameter $A$ according to $A \leq A_\textrm{max} = 1.3 \, x^{0.33}$ and 
the maximum value of $\xi/\xi_0$, which depends on the combination $A\, x \leq 1.3 \, x^{1.33}$. %, with $x= \tau/\xi_0$. %, as shown in Figure \ref{xsweep}. 
Taking the more conservative, but still optimistic value of $\tau = 1$ fm yields, for $\xi_0 = 1.6$ fm, a maximum value of $\xi/\xi_0 = 1.3$, instead 
of the previous maximum of $\xi/\xi_0 = 1.9$. 

Apart from the different value of $\tau$, the treatment described here is the same as in Ref.~\cite{Hippert2016}, where some of its aspects are discussed in more 
detail and the distribution~(\ref{eqPsigma02}) is used to build  a simple Monte Carlo implementation of critical fluctuations to which background 
contributions can be superposed.

\subsection{Background fluctuations}
\label{secBGf}

One of the most challenging limitations in the search for fluctuations which might serve as a probe of criticality is understanding the 
role and behavior of background contributions. These contributions may come from any source of undesired statistical correlation, including 
quantum statistics, fluctuations of imperfectly controlled thermodynamic parameters, such as temperature and system size, and 
correlations from initial conditions~\cite{Stephanov:1999zu,Hippert2016,Luo:2017faz,Gorenstein2015,Sahoo2013,Herold2016,Luo2013a}. 
An experimental discovery of the critical point can hardly be claimed before background is either controlled, well described in terms of 
beam energy dependence or eliminated by a good choice of signatures. 

We use the same model for spurious fluctuations as in Ref.~\cite{Hippert2016}. For temperature fluctuations we simply take a Gaussian distribution of $5\%$ width. 
For volume fluctuations, we take a probability distribution $\mathcal{P}_b (b) \propto b$ for the impact parameter $b$, and suppose the final volume $V$ to be 
proportional to the initial overlap area $\mathcal A$ between the two nuclear disks of radius $R_{N}$, with a proportionality constant $C$,
\begin{align}
\begin{split}
V(b, R_{N}) &= C\, \mathcal A(b, R_{N}) \\
 &= 2\, C\,  R_{N}^2 \cos^{-1}\left(\dfrac{b}{2 R_{N}} \right) - b \sqrt{R_{N}^2 - \dfrac{b^2}{4}}\, , 
\end{split}
\end{align}
in which we take $R_{N} = r_0 = 6.38$~fm for gold nuclei (see Refs.~\cite{DeVries1987495,DeJager1974479}) and $C$ can be fixed by the 
average radius, or average volume, of the system at freeze-out. 
This model enables us to find the system-volume distribution in a given centrality range, while disregarding initial-state fluctuations and other effects \cite{Luo2013a}.

These are very crude models for background fluctuations, but more complete models can be very straightforwardly included within the present framework. 
Volume fluctuations have previously been considered in Refs.~\cite{Luo:2017faz,Skokov2012,Gorenstein2011,Luo2013a}, as well as the methods for partially correcting them in data analysis~\cite{Luo:2017faz,Luo2013a}.  
Certain fluctuation measures, called strongly intensive, are, by construction, insensitive to such fluctuations~\cite{Gorenstein2011,Gorenstein2015}.

\section{Computing fluctuations from energy-level shifts}%Analytical approach
\label{secMethod}

\subsection{Methodology}

Besides simulating fluctuations using Monte Carlo techniques~\cite{Hippert2016}, it is possible to derive analytical expressions for signatures of the critical point 
within this scenario. 
We now develop a scheme to calculate observables in the presence of spurious and critical fluctuations. It has the advantage of naturally including finite-size effects. 

Inspired by Sec.~\ref{secCritFluct}, we note that a fluctuation of the order parameter $\sigma_0$, produces a  shift in 
each single-particle energy level $\omega$,
\begin{align}
\begin{split}
  \omega & =  \sqrt{p^2 + m_0^2 + \delta m^2 } \\
      & \approx   \omega_0\left[1+\dfrac{1}{2} \dfrac{\delta m^2}{\omega_0^2} - \dfrac{1}{8} \dfrac{(\delta m^2)^2}{\omega_0^4} + \cdots \right]\\ 
       & \approx  \omega_0\left[1+ \dfrac{m \, \delta m}{\omega_0^2} + \dfrac{1}{2 }\left( 1 - \dfrac{m^2}{\omega_0^2}\right)\dfrac{( \delta m)^2}{\omega_0^2}  + \cdots \right]\,,
\label{dwds}
\end{split}
\end{align}
where $\delta m = g\; \delta \sigma$ is the corresponding mass correction and $\omega_0$ is the original one-particle energy.

Spurious fluctuations of parameters such as volume, temperature and chemical potential can also be described as energy-level shifts. 
Suppose we have a gas of free particles confined to a typical length scale $R$. 
For both cubic and spherically symmetric boundary conditions, the allowed momentum levels will be distributed depending on the value of $R$, 
$ p_{i} = {\alpha_{i}}/{R}$, where $\alpha_{i}$ represents constants which depend on the boundary conditions.  
A small change $\delta R$ in the size of the system, hence, modifies the momenta, $p_i$, and the one-particle energies, $\omega_i$, of the allowed modes,
\begin{align}
 p_i= & \dfrac{\alpha_{i}}{R +\delta R} \approx p_{0\, i} \left[ 1 - \dfrac{\delta R}{R} + \left(\dfrac{\delta R}{R}\right)^2 + \cdots\right] \,,\\ %+ {\cal O}(\left({\delta R}/{R}\right)^3)
\begin{split} 
\omega_i= & \sqrt{p_i^2 + m^2} \approx \omega_{0\,i} \left\{ 1-\left(\dfrac{p_{0\,i}}{\omega_{0\,i}}\right)^2 \dfrac{\delta R}{R} + \right. \\
          &\left. + \dfrac{1}{2} \left[ 3\left(\dfrac{p_{0\,i}}{\omega_{0\,i}}\right)^2 -\left(\dfrac{p_{0\,i}}{\omega_{0\,i}}\right)^4 \right] \left(\dfrac{\delta R}{R}\right)^2  + \cdots\right\} \,. 
\label{dwdR}
\end{split}
\end{align}

Temperature and chemical potential fluctuations can be included in a similar fashion. 
All the quantities we calculate in this work depend on temperature, chemical potential and energy levels only through the Boltzmann factor 
$e^{-(\omega_i-\mu)/T}$. As a consequence, fluctuations of both temperature and chemical potential are equivalent to fluctuations of the energy levels. 
For a fluctuation $\delta T$ in the temperature, for instance,
\begin{equation}
 \dfrac{\omega_{0\,i} - \mu}{T + \delta T} =  \dfrac{\omega_{0\,i} - \mu}{T}\left(1 - \dfrac{\delta T}{T} + \dfrac{\delta T^2}{T^2} + \cdots \right)\,,
\end{equation}
so it is equivalent to a fluctuation in $\omega_i$ of the form
\begin{equation}
 \delta \omega_i^T = (\omega_{0\,i} - \mu)\left(- \dfrac{\delta T}{T} + \dfrac{\delta T^2}{T^2} + \cdots \right)\,.
 \label{dwdT}
\end{equation}
For a fluctuation $\delta \mu$ in the chemical potential,
\begin{equation}
 \omega_{0\,i} - (\mu + \delta \mu) = \omega_{0\,i} + \delta\omega^\mu_i - \mu\,,
 \label{dwdmu}
\end{equation}
with $\delta\omega^\mu_i = - \delta \mu$.

In order to calculate the effect of these fluctuating energy shifts, an ensemble of different $(\delta \sigma, \delta R, \delta T, \delta \mu, \cdots)$ and occupation numbers, $\{n_i\}$, can be considered. 
Furthermore, configurations in this ensemble can be organized according to the values of $\delta R$ and $\delta \sigma$, so that averages can be calculated by first averaging 
over a regular grand-canonical ensemble for a fixed  $(\delta \sigma, \delta R, \delta T, \delta \mu, \cdots)$ (denoted by $\langle \cdots \rangle$) and then averaging over  
$(\delta \sigma, \delta R, \delta T, \delta \mu, \cdots)$ (denoted by $\overline{\cdots}$).

We proceed by expanding the first averages, $\langle \cdots \rangle$, as  power series in $\delta \sigma, \delta R, \cdots$ before averaging over their different values. 
Defining $\overline{\delta \sigma } =\overline{\delta R} = \overline{\delta T} =\overline{\delta \mu} = 0$, 
an expansion to at least order $(\delta \omega)^2$ in the single-particle energy corrections is necessary for consistency.

\subsection{Second-order expansion}
\label{secAnO2}

We now turn to the actual calculations. 
The lowest-order contributions to observables come from the variance of each fluctuating quantity, $\overline{\delta \sigma^2}$,  $\overline{\delta R^2}$ and so on, 
corresponding to a Gaussian approximation. 
Before we go further, it is useful to introduce some relations, which can be derived from $\partial^n (\beta F)/\partial \zeta_i^n = \langle (\Delta n_i)^n \rangle$:
\begin{align}
\begin{split}
&\langle n_i \rangle = f(\zeta_i) :=  (e^{-\zeta_i} \mp 1)^{-1} \,, \\
&\langle(\Delta n_i)^2 \rangle = f^\prime(\zeta_i) =   f(\zeta_i) \;\big(1 \pm f(\zeta_i)\big) \,, \\
&\langle(\Delta n_i)^3 \rangle = f^{\prime\prime}(\zeta_i) = f^\prime(\zeta_i)\;\big(1\pm 2\, f(\zeta_i)\big)\,,\\
&\langle(\Delta n_i)^4 \rangle = f^{\prime\prime\prime}(\zeta_i) = f^\prime(\zeta_i)\big(1\pm 6\, f^\prime(\zeta_i)\big) \,,
\label{cumu_n-free}
\end{split}
\end{align}
where $\Delta n_i := n_i - \langle n_i \rangle$, $F$ is the free-energy, $\beta := 1/T$, $\zeta_i := - \beta (\omega_i- \mu)$ and the upper (lower) sign stands for bosons (fermions).

We start by the average occupation number of the $i$th~energy level. 
Expanding the ordinary expression for the grand-canonical ensemble up to $\delta \omega_i^2$, for noninteracting particles, 
\begin{align}
\begin{split}
{\langle n_i \rangle} \approx & f(\zeta_{0\,i}) - \beta \; f^\prime(\zeta_{0\,i}) \;{\delta \omega_i}  + \\ 
 & + \dfrac{\beta^2}{2} f^{\prime\prime}(\zeta_{0\,i}) \;{(\delta \omega_i)^2} + \cdots \,,
\end{split}
\end{align}
and taking the average over  fluctuations of the induced shifts, 
\begin{align}
\begin{split}
 \overline{\langle n_i \rangle} \approx & f(\zeta_{0\,i}) - \beta \; f^\prime(\zeta_{0\,i}) \; \overline{\delta \omega_i}  + \\ 
 & + \dfrac{\beta^2}{2} f^{\prime\prime}(\zeta_{0\,i}) \;\overline{(\delta \omega_i)^2} + \cdots \,,
\label{avn1}
\end{split}
\end{align}
where $\zeta_{0\,i} : = - \beta (\omega_{0\,i}- \mu_0)$. 

A calculation of the microscopic 
correlator\footnote{When in an overlined expression, $\Delta n_i := n_i - \overline{\langle n_i \rangle}$.}~\cite{Stephanov:1999zu}
is also possible,
\begin{align}
\begin{split}
 \overline{\langle \Delta n_i\, \Delta n_j \rangle} :=& \overline{\langle n_i n_j \rangle} - \overline{\langle n_i \rangle} \; \overline{\langle n_j \rangle}\\
=& \overline{\langle n_i \rangle \langle n_j \rangle}  - \overline{\langle n_i \rangle} \; \overline{\langle n_j \rangle} + \delta_{i j} \, \overline{f^{\prime}(-\beta \omega_i)}\,.
\end{split}
\end{align}
From Eq.~(\ref{avn1}), since $\overline{\delta \omega_i} \,\overline{\delta \omega_j} \sim {\cal O} (\delta \sigma^4, \delta R^4, \cdots)$,
\begin{multline}
  \overline{\langle \Delta n_i\, \Delta n_j \rangle} \approx f^\prime(\zeta_{0\,i})\,  f^\prime(\zeta_{0\,j})\; \beta^2 \,  \overline{\delta\omega_i \delta \omega_j} + \\
+\delta_{i j} \bigg( f^\prime(\zeta_{0\,i})- \beta \; f^{\prime\prime}(\zeta_{0\,i}) \;\overline{\delta \omega_i} + \\
+\dfrac{\beta^2}{2} f^{\prime\prime\prime}(\zeta_{0\,i}) \;\overline{(\delta \omega_i)^2} \bigg)\,,
\label{micCor1}
\end{multline}
where  derivatives of $f(\zeta_i)$ can be extracted from Eq.~(\ref{cumu_n-free}) and we have used that $\langle \Delta n_i \Delta n_j \rangle = \delta_{i j} \, \langle (\Delta n_i)^2 \rangle$ for free particles.

Finally, from Eqs.~(\ref{dwds}), (\ref{dwdR}), (\ref{dwdT}) and (\ref{dwdmu}), neglecting terms of order $(\delta \sigma\, \delta R)^2$, for instance, and assuming 
independent fluctuations,\footnote{It is, of course, very straightforward to relax this assumption, provided the relevant correlations.} 
\begin{multline}
   \overline{\delta \omega_i} \approx   \dfrac{\omega_{0\,i}}{2} \left[ 3\left(\dfrac{p_{0\,i}}{\omega_{0\,i}}\right)^2 
  -\left(\dfrac{p_{0\,i}}{\omega_{0\,i}}\right)^4 \right]\, \overline{\left(\dfrac{\delta R}{R}\right)^2} + \\
+  \dfrac{g^2 }{2\, \omega_{0\,i}}\left( 1 - \dfrac{m^2}{\omega_{0\,i}^2}\right)\, \overline{( \delta \sigma)^2} 
+ (\omega_{0\,i} - \mu)\, \overline{\left(\dfrac{\delta T}{T}\right)^2}\,,
\label{avomega}
\end{multline}
\begin{multline}
\overline{\delta \omega_i\, \delta \omega_j} \approx \dfrac{g^2\, m^2}{\omega_{0\,i}\, \omega_{0\,j}}\, \overline{( \delta \sigma)^2} - 
\dfrac{(p_{0\,i}\, p_{0\,j})^2}{\omega_{0\,i}\, \omega_{0\,j}} \,\overline{\left(\dfrac{\delta R}{R}\right)^2} \\
+ (\omega_{0\,i} - \mu)(\omega_{0\,j} - \mu)\, \overline{\left(\dfrac{\delta T}{T}\right)^2} + \overline{(\delta \mu)^2}\,.
\label{avomega2}
\end{multline}

The average and variance of the particle multiplicity can then be written as
\begin{equation}
 \overline{\langle N \rangle} = \sum_{i}  \overline{\langle n_i \rangle}\,,
\end{equation}
\begin{equation}
 \overline{\langle (\Delta N)^2 \rangle} = \sum_{i,j}  \overline{\langle \Delta n_i\, \Delta n_j \rangle}\,.
\end{equation}

The results presented above can be generalized to other observables and higher-order moments, by including higher orders of $\delta \omega$. 
Since volume fluctuations modify the momentum modes,  the calculation of fluctuation measures 
involving the momenta requires the computation of terms such as $\overline{p_i \, \langle \Delta n_i\, \Delta n_j \rangle}$ and 
$\overline{p_i\, p_j\, \langle \Delta n_i\, \Delta n_j \rangle}$, for instance.

\subsection{Hard-sphere boundary conditions}
\label{secBC}

The finite size of the system 
requires the specification of boundary conditions in order to use results from Sec.~\ref{secAnO2}.
We choose to work with Dirichlet boundary conditions on a sphere of radius $R$,
\begin{equation}
\Phi(R,\theta, \phi) = 0\,. 
\end{equation}
 Although they are not very realistic, these boundary conditions are more natural than cubic ones and yet fairly simple. 
 
  The eigenstates of the squared momentum $p^2$ in a spherical geometry are given by the spherical Bessel functions of the first kind $j_\ell (p\,r)$, 
 where the orbital angular momentum $\ell$ is a positive  integer.  The allowed eigenvalues of $p$ are given by $p^{(\ell)}_i= \alpha_i^{(\ell)}/R$, where $\alpha^{(\ell)}_i$ 
 is the $i$th~root of $j_\ell (x)$, $j_\ell (\alpha^{(\ell)}_i) = 0$. For each pair $(\ell,i)$ there are $2\,\ell+1$ linearly independent  one-particle quantum states, 
 corresponding to eigenvalues of the $z$ component of the angular momentum $-\ell\leq m \leq \ell$.

In the absence of kinematic cuts,
\begin{align}
\begin{split}
  \overline{\langle (\Delta N)^2 \rangle} =& \sum_{\substack{\ell_1,m_1,i_1\\ \ell_2,m_2,i_2}}\overline{\langle \Delta n_{\ell_1, i_1}\, \Delta n_{\ell_2, i_2} \rangle }\\
 =& \sum_{\substack{\ell_1, i_1\\\ell_2,i_2}}  (2\,\ell_1+1) (2\,\ell_2+1)\; A_{(\ell_1, i_1),(\ell_2,i_2)} + \\
 & + \sum_{\ell_1, i_1}  (2\,\ell+1)\;B_{(\ell,i)}\,,
 \label{micCorBC}
\end{split}
\end{align}
where, from Eq.~(\ref{micCor1}),
\begin{multline}
 A_{(\ell_1, i_1),(\ell_2,i_2)} = f^\prime(\zeta_{0\,(\ell_1, i_1)})\,  f^\prime(\zeta_{0\,(\ell_2,i_2)})\times \\
\times\beta^2 \,  \overline{\delta\omega_{(\ell_1, i_1)} \delta \omega_{(\ell_2,i_2)}}\,,
\label{corrcoefA}
\end{multline}
\begin{multline}
   B_{(\ell,i)}  =  f^\prime(\zeta_{0\,(\ell_1, i_1)})-\beta \; f^{\prime\prime}(\zeta_{0\,(\ell_1, i_1)}) \;\overline{\delta \omega_{(\ell_1, i_1)}} + \\
+\dfrac{\beta^2}{2} f^{\prime\prime\prime}(\zeta_{0\,(\ell_1, i_1)}) \;\overline{(\delta \omega_{(\ell_1, i_1)})^2}\,.
\label{corrcoefB}
\end{multline}

It is also possible to calculate moments involving different kinds of particles, $A$ and $B$, such as $\overline{\langle \Delta N^A \Delta N^B \rangle}$, for instance. 
In this case, the previous calculations apply with small modifications and we have, in Eq.~(\ref{micCorBC}),
\begin{multline}
 A_{(\ell_1, i_1),(\ell_2,i_2)} = f^\prime_A(\zeta^A_{0\,(\ell_1, i_1)})\,  f^\prime_B(\zeta^A_{0\,(\ell_2,i_2)})\times \\
\times\beta^2 \,  \overline{\delta\omega^A_{(\ell_1, i_1)} \delta \omega^B_{(\ell_2,i_2)}}\,,
\end{multline}
\begin{equation}
   B_{(\ell,i)}  =  0\,,
\end{equation}
where Eq.~(\ref{avomega2}) can be used with adaptations for the chemical potentials, masses and couplings.

\section{Acceptance cuts}

\label{secKC}

Expression~(\ref{micCorBC}) is valid for the total multiplicity of particles under the chosen 
boundary conditions. 
 Experimentally, however, only the particles produced in a  certain range 
of transverse momentum and pseudorapidity are detected and used in data analysis.  
The behavior of fluctuation measures under variations of these acceptance cuts has been 
discussed, for example, in Refs.~\cite{Ling2016,Karsch2015,Bzdak2012,Garg2013,PhysRevC.66.044904}. 
Here we show how they can be included in our framework. 

The quantum numbers $\ell$ and $p$, chosen in Sec.~\ref{secBC}, give us no information about the separate longitudinal and transverse parts of the momentum.  
However, it is possible to calculate the fraction of particles of a given momentum $p$ which will be in the selected window of rapidity and transverse momentum, 
assuming their momenta  are distributed isotropically and without angular correlation. 
Because of quantum statistics, this assumption is not obviously true 
for indistinguishable particles, but should hold as a reasonable approximation.

The upper and lower transverse momentum cuts,  $p_T \leq p_{+}$ and $p_T \geq p_{-}$, may be written as 
\begin{equation}
\begin{split}
 u^2 \geq 1 - \left(\dfrac{p_{+}}{p} \right)^2\,,\;\;\;\;
\end{split}
\begin{split}
   u^2 \leq 1 - \left(\dfrac{p_{-}}{p} \right)^2\,.
\end{split}
\end{equation}
where $u:=\cos \theta$, $p_T := p \, |\sin \theta|$ and $\theta$ is the angle between the momentum of the particle 
and the beam axis.

We also select a window for the pseudorapidity $\eta = \ln[(p+p_z)/(p-p_z)]/2$. 
Since $\eta = \ln[(1+ u)/(1-u)]/2$, $|\eta| \leq \eta_C$ yields
\begin{equation}
e^{-2\,\eta_C} \leq \dfrac{1+u}{1-u} \leq e^{2\,\eta_C}\,,
\end{equation}
resulting in 
\begin{equation}
 e^{-2\,\eta_C} -1 \leq u (1 + e^{-2\,\eta_C}) \;\;\;\;\; u (1+ e^{2\,\eta_C}) \leq e^{2\,\eta_C}-1\,,
\end{equation}
\begin{equation}
 |u| \leq \tanh \eta_C\,.
\end{equation}

\begin{figure}
\includegraphics{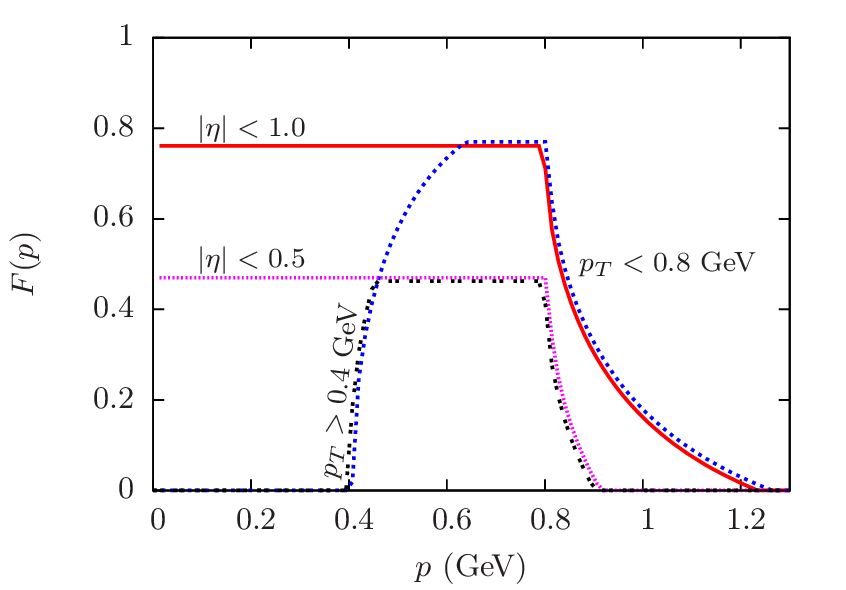}
	\caption{Particle acceptance probability or fraction as a function of its momentum $p$ for isotropic emission. 
We display curves corresponding to $|\eta|<1.0$, $|\eta|<0.5$ and $p_T>0$, $p_T>0.4$ GeV, with  
$p_T<0.8$ GeV for all of the curves. Lines were slightly shifted to avoid superposition.}
\label{figAcc-F}
\end{figure}

Combining both rapidity and transverse momentum cuts, we have 
\begin{equation}
\max\left[0, {\frac{p^2-p_+^2}{p^2}}\right] \leq u^2 \leq \min\left[{\frac{p^2 -p_-^2}{p^2}}, \tanh^2 \eta_C\right]\,.
\label{rangeu2}
\end{equation}
Since we assume an uncorrelated and uniform angular distribution of particles and $|u|$ is uniformly distributed in the interval $(0,1]$, 
the resulting fraction of accepted particles is 
\begin{equation}
 F(p) = \int_{\Omega_{\textrm{acc}}(p)}\dfrac{d\Omega}{4\,\pi}=\max[u_{\rm{max}}(p) - u_{\rm{min}}(p), 0]\,,
\end{equation}
where
 \begin{align}
u_{\rm{max}}(p) &= 
 \begin{cases}
 \min\left[{\sqrt{\frac{p^2 -p_-^2}{p^2}}}, \tanh \eta_C\right] & p > p_-\\
 0 & p < p_-
  \end{cases}
  \,,\\
 u_{\rm{min}}(p) &= 
 \begin{cases}
  \sqrt{\frac{p^2-p_+^2}{p^2}} & p > p_+\\
  0 & p < p_+
 \end{cases}
 \,.
\end{align}
A plot of $F(p)$ for different acceptance windows is shown in Fig.~\ref{figAcc-F}.

Event-by-event cumulants of observables may now be calculated. 
Attention should be drawn to the fact that the acceptance of particles 
is probabilistic  and follows a binomial distribution,  
which also contributes to fluctuations. 
Hence, denoting as $\langle \cdots \rangle_{\textrm{acc}}$ an average over accepted particles only, for a single value of $p$, 
\begin{equation}
  \langle n_p \rangle_{\textrm{acc}} = F(p) \, \langle n_p \rangle\,,
\end{equation}
\begin{equation}
 \langle  n_p^2 \rangle_{\textrm{acc}} = \langle n_p\,F(p) + n_p (n_p - 1)\, F(p)^2 \rangle\,,
\end{equation}
with which,  
\begin{multline}
\overline{\langle (\Delta n_p)^2 \rangle}_{\textrm{acc}}= \overline{F(p)^2 \, \langle n_p^2 \rangle} - \left(\overline{F(p)\, \langle n_p \rangle}\right)^2 +\\
+ \overline{F(p)\big(1 - F(p)\big)\, \langle n_p \rangle}\,,
\end{multline}
where we denote $\Delta n_p := n_p - \langle n_p \rangle_{\textrm{acc}}$ or $\Delta n_p := n_p - \overline{\langle n_p \rangle}_{\textrm{acc}}$ in the 
corresponding averages over accepted particles. 

For the sake of simplicity, we neglect fluctuations of $F(p)$. 
For midrapidity, $F(p)$ is only sensitive to $p$ in the limits of the acceptance region, as seen in Fig.~\ref{figAcc-F}, 
and fluctuations of $p$ should only be relevant for the acceptance probability in these narrow regions.   
In this approximation, 
\begin{multline}
 \overline{\langle (\Delta n_p)^2 \rangle}_{\textrm{acc}}= F(p_0)^2 \, \overline{\langle (\Delta n_p)^2 \rangle} +
\\+ F(p_0)\big(1 - F(p_0)\big)\, \overline{\langle n_p \rangle} \,,
\label{approx-kc-1p}
\end{multline}
where $p_0$ is the unperturbed value of the momentum mode in question. 
A similar result is found for the correlator of different momentum modes, while lacking the second term in Eq.~(\ref{approx-kc-1p}). 

Hence, Eq.~(\ref{micCorBC}) is modified to 
\begin{multline}
   \overline{\langle (\Delta N)^2 \rangle} = \sum_{\substack{l_1,m_1,i_1\\l_2,m_2,i_2}}\overline{\langle \Delta n_{\ell_1, i_1}\, \Delta n_{l_2, i_2} \rangle }_{\textrm{acc}} \\
=\sum_{\substack{\ell_1, i_1\\\ell_2,i_2}}  (2\,\ell_1+1)\, F(p_{i_1}^{(l_1)}) \, (2\,\ell_2+1)\, F(p_{i_2}^{(l_2)})\, A_{(\ell_1, i_1),(\ell_2,i_2)}+\\
+ \sum_{\ell_1, i_1}  (2\,\ell+1)\; \bigg[ F(p_i^{(\ell)})^2\, B_{(\ell,i)} + 
\\ + F(p_i^{(\ell)})\big(1 - F(p_i^{(\ell)})\big)\, C_{(\ell,i)} \bigg]\,,
\label{varNKC}
\end{multline}
where $A_{(\ell_1, i_1),(\ell_2,i_2)}$ and $B_{(\ell,i)}$ are given by Eqs.~(\ref{corrcoefA}) and (\ref{corrcoefB}) and $C_{(\ell,i)}$ is found from Eq.~(\ref{avn1}),
\begin{multline}
C_{(\ell,i)} = f(\zeta_{0\,(\ell, i)}) - \beta \; f^\prime(\zeta_{0\,(\ell, i)}) \; \overline{\delta \omega_{(\ell, i)}}  + \\ 
 + \dfrac{\beta^2}{2} f^{\prime\prime}(\zeta_{0\,(\ell, i)}) \;\overline{(\delta \omega_{(\ell, i)})^2} + \cdots \,.
\label{coefCvarNKC}
\end{multline}

As in Ref.~\cite{Bzdak2012}, we use a probabilistic description for acceptance effects. 
However, we take into account the momentum dependence in the relevant probabilities. 
A similar treatment will be applied to resonance decay under acceptance cuts in the next section. 
Note that a finite detection efficiency $e_0(p)$ can be introduced by making $F(p) \rightarrow e_0(p) \cdot F(p)$. 

The implementation of kinematic cuts in our Monte Carlo simulations is far simpler and can be done by 
independently sampling the momentum direction for each of the produced particles and applying the cuts, 
also relying on the assumption of uncorrelated, isotropic emission. 
It is also straightforward to include the effects of a finite efficiency. 

\section{Resonance decay effects}
\label{secRes}

So far, we have only considered direct particles from a thermal distribution. 
However, a non-negligible fraction of the final-state particles in a collision come from the decay of unstable particles. 
In this section, we consider the role of resonance decay in multiplicity fluctuations. 
The impact of resonance decay in fluctuation measures was previously explored in Refs.~\cite{Bluhm:2016byc,Mishra2016,Nahrgang2015,Sahoo2013,Garg2013,Begun2006,Jeon1999}, 
among others. 

We consider particles coming from resonance decay to be affected by spurious fluctuations, but 
not by critical ones, such that they are expected to dilute signatures of criticality. 
Moreover, both the decay of unstable particles and
the detection of its products in the relevant acceptance window are probabilistic processes, 
which affects fluctuations of the particle multiplicities. 

Resonance decay contributions to particle multiplicity fluctuations can be incorporated to 
our calculations by using the probability of having each decay product 
in the acceptance window. These probabilities are computed in Sec.~\ref{secprobResDec} 
and later used in Sec.~\ref{secFluctResDec} to calculate the desired contributions. 
Unlike most previous calculations of resonance decay effects, we apply acceptance cuts to the decay products 
 themselves,  not to the resonances. 
For simplicity, we neglect the widths of the resonances. 
As an example, we apply our methods to the decay of a $\rho$ meson into two pions.

\subsection{Probabilistic treatment}
\label{secprobResDec}

We aim to calculate the distribution of decay products found within the acceptance window. 
Information about both the average fraction of accepted particles  and the corresponding fluctuations is necessary. 
We limit ourselves to two-particle decays and start with branching ratios of $100\%$. 
We use the associated phase-space volume as a measure of probability for the kinematic variables. 

We denote the momentum of the  resonance by ${\bf p}_{in}$.  Since we assume spherical symmetry, there is no preferred direction for it. 
The products of the decay have momenta ${\bf p}_1$ and ${\bf p}_2$, with orientations given by polar and azimuthal angles of $\theta_1$, $\theta_2$ and $\phi_1$, $\phi_2$ , respectively. 
Imposing energy and momentum conservation, the two-particle phase-space volume differential reads
\begin{multline}
 d \Phi_{2 P} \sim \dfrac{p_1^2 \, d p_1 \, d\Omega_1}{2\, E_{1}} \; \dfrac{p_2^2 \, d p_2 \, d\Omega_2}{2\, E_{2}} \; \delta(E_1 + E_2 - E_{in})\times \\
\times \; \int_{4\pi} d\Omega_{in} \;\delta^{(3)}({\bf p}_1 + {\bf p}_2 - {\bf p}_{in})\,,
\end{multline}
where $E_i := \sqrt{p^2_i + m^2_i}$ and we integrate the direction of the resonance momentum over the whole sphere.

Changing variables from $\theta_2, \phi_2$ to $\theta^{(1)}_2, \phi^{(1)}_2$, defined with their zenith in ${\bf p}_1/p_1$,
\begin{multline}
 \int_{4\pi} d\Omega_{in} \;\delta^{(3)}({\bf p}_1 + {\bf p}_2 - {\bf p}_{in}) = \dfrac{1}{p^2_{in}} \delta(|{\bf p}_1 + {\bf p}_2|-p_{in})\\
 = \dfrac{1}{p_{in}\, p_1\, p_2}\delta(\cos \theta^{(1)}_2 - \cos \theta^*)
\end{multline}
with $\cos \theta^* :=({p_{in}^2 - p_1^2 - p_2^2})/2\, p_1  p_2$. 
Hence, using $dE_i/dp_i = p_i/E_i$,
\begin{multline}
 d \Phi_{2 P} \sim  \dfrac{1}{p_{in}} \;d\phi^{(1)}_2\, d\phi_1\, d(\cos \theta_1)  \, d E_1 \times \\
  \times d E_2 \; \delta(E_2 - E_{in}+ E_1)\times \\ 
\times  d(\cos\theta^{(1)}_2) \; \delta(\cos \theta^{(1)}_2 - \cos \theta^*)\,.
\label{eqPS2}
\end{multline}
If we interpret $d \Phi_{2 P}$ as proportional to a probability density, Eq.~(\ref{eqPS2}) suggests $\phi^{(1)}_2$, $\phi_1$, $\cos \theta_1$ and $E_1$ 
to be uniformly distributed among their full range of values, while $E_2$ and $\theta^{(1)}_2$, the angle between ${\bf p}_1$ and ${\bf p}_2$, 
are determined by the former. 
Notice that the conditions that $E_2^* := E_{in}- E_1$ and $\cos \theta^*$ fall within the integration regions for $E_2$ and $\cos \theta^{(1)}_2$ 
implicitly limit the possible values of $E_1$. 
The constraint $\cos^2 \theta^*|_{p_2=p_2^*} \leq 1$, for example, yields
\begin{equation}
 E_0 - \dfrac{1}{2} \Delta E \leq  E_1 \leq  E_0 + \dfrac{1}{2} \Delta E\,,
\label{eqCutEres}
\end{equation}
with 
\begin{equation}
 E_0 :=  \dfrac{m_{in}^2 + m_1^2 - m_2^2}{m_{in}^2}\, \dfrac{E_{in}}{2}\,,
\end{equation}
\begin{equation}
 \Delta E := \dfrac{p_{in}}{m_{in}^2}\sqrt{(m_{in}^2 - m_1^2 - m_2^2)^2 - 4\, m_1^2\, m_2^2}\,.
\end{equation}
% For $m_1=m_2=140$ MeV and $m_{in}=770$ MeV, the  

 \begin{figure}
  \includegraphics{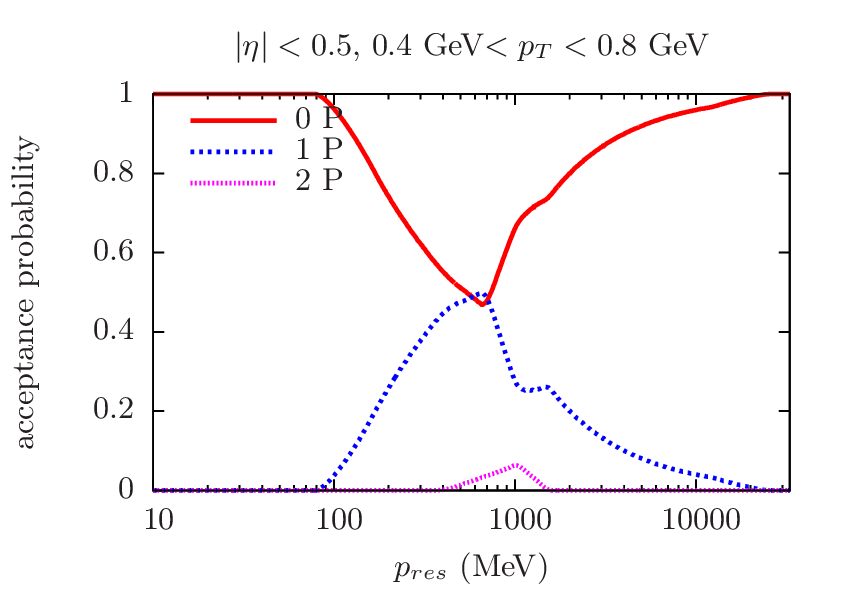}
\caption{Probability of accepting $0$, $1$ or $2$ of the two identical particles of mass $m=140$~MeV coming from the decay of a resonance of mass $770$~MeV, 
with kinematic cuts $0.4$~GeV$< p_T <0.8$~GeV, $|\eta|< 0.5$.}
\label{ProbAcc}
 \end{figure}

It is now possible to sample the number of decay products inside the acceptance window for a resonance of momentum $p_{in}$. 
We sample $\phi^{(1)}_2$, $\phi_1$, $\cos \theta_1$ and $E_1$ uniformly and calculate $E_2$ and $\cos \theta^{(1)}_2$, counting 
the rate of acceptance within kinematic cuts to find the probability of accepting each particle.\footnote{The
change of variables  $\theta_2, \phi_2 \leftrightarrows \theta^{(1)}_2, \phi^{(1)}_2$ can be undone by a simple rotation.} 
Results are exemplified in Fig.~\ref{ProbAcc}. 
Condition~(\ref{eqCutEres}), together with the form of $F(p)$ (see Fig.~\ref{figAcc-F}), is responsible 
for the  acceptance probabilities in Fig.~\ref{ProbAcc} vanishing both above $p_{res} \approx 25$~GeV and bellow 
 $p_{res} \approx 80$~MeV.\footnote{We thank V. Koch for pointing out that these probabilities should vanish at high $p_{res}$.} 

In Monte Carlo simulations, the momenta of the decay products can be sampled according to the above scheme 
for each single resonance. 

\subsection{Fluctuations from resonance decays}
\label{secFluctResDec}

We now apply the results from Sec.~\ref{secprobResDec} in the context of particle multiplicity fluctuations. 
 Consider the decay of a resonance $R$ into two distinct particles, $A$ and $B$, with probabilities $P_A$, $P_B$, $P_2$ and $P_0$ that only 
 particle $A$, only particle $B$, both particles or neither of them are within the acceptance window, respectively. 
These probabilities provide all the relevant information for the study of  multiplicity fluctuations. 
Then, we calculate averages involving the number of accepted particles, $n_l$, where $l=A,B$. 
For the decay of a single resonance with momentum $p$, for instance, $n_l$ is either $0$ or $1$ and\footnote{For simplicity 
of notation, we choose to hide the dependence on  $p$ of $P_l$, $P_2$, $P_0$ and kinematic averages.} 
\begin{equation}
 \langle (n_l)^m \rangle_{\textrm{kin}} = P_l + P_2\,,
\end{equation}
\begin{equation}
\langle (n_A\,n_B)^m\rangle_{\textrm{kin}} = P_2\,,
\end{equation}
where $m \in \mathbb{N}$ and $\langle \cdots \rangle_{\textrm{kin}}$ denotes an average over the angular distribution 
of particles within kinematic cuts. 
For two different decays, $i$ and $j$, of resonances with the same momentum $p$,
\begin{equation}
 \langle n_{l}^{(i)} \, n_{l^\prime}^{(j)} \rangle_{\textrm{kin}} = (P_l + P_2)(P_{l^\prime} + P_2).
\end{equation}

Considering now the independent decays of $n^R_p$ resonances of momentum $p$, summing over $p$ and taking averages over 
thermodynamic, critical and spurious fluctuations yields  
for the contribution from 
 correlations among resonance decay products: 
\begin{equation}
 \overline{\langle n_l \rangle}_R = \sum_{p}\, \overline{\langle n^R_p\rangle}\, (P_l + P_2 )\,,
\end{equation}
\begin{multline}
 \overline{\langle (\Delta n_l)^2\rangle}_{R,R} =  \sum_{p, p^\prime} \overline{\langle \Delta n^R_p\, \Delta n^R_{p^\prime}\rangle} (P_l + P_2)(P^\prime_l + P^\prime_2) + \\ 
+ \sum_p \overline{\langle n^R_ p \rangle} (P_l + P_2) (1 - P_l - P_2) \,.
\label{diagResCor}
\end{multline}
\begin{multline}
 \overline{\langle \Delta n_A \, \Delta n_B\rangle}_{R,R} =  \sum_{p, p^\prime} \overline{\langle \Delta n^R_p\, \Delta n^R_{p^\prime}\rangle} (P_A + P_2)(P^\prime_B + P^\prime_2) + \\ 
+ \sum_p \overline{\langle n^R_ p \rangle} \big[P_2 - (P_A + P_2)(P_B + P_2)\big]\,,
\label{cruzadoResCor}
\end{multline}
As in Sec.~\ref{secKC}, we neglect the influence of fluctuations on the fraction of accepted particles,  
taking the acceptance probability for the unperturbed resonance momentum.\footnote{In this case,
however, these probabilities are also affected by changes in the mass of the particles, driven by critical fluctuations.} 
Such effects are present in the full simulations. 
For the contributions from correlations between resonance decay products and direct particles:
\begin{multline}
 \overline{\langle\Delta n_l \, \Delta n_{l^\prime}\rangle}_{R,\textrm{dir}} =  \sum_{p}\, \overline{\langle \Delta n^R_p \Delta n_{l^\prime}\rangle}\, (P_l + P_2 )\,.
\end{multline}
Contributions from correlations between decays of different resonances can also be calculated in a straightforward fashion.

A similar treatment can be applied to decays into two identical particles, with $P_A=P_B=P_1$, in which case, 
for a single decay, $n_l$ is either $0$, $1$ or $2$ and 
\begin{equation}
 \langle (n_A)^m \rangle_{\textrm{kin}} = P_1 + 2^m\,P_2\,,
\end{equation}
\begin{equation}
 \overline{\langle n_A \rangle}_R = \sum_{p}\, \overline{\langle n^R_p\rangle}\, (P_1 + 2\,P_2 )\,,
\end{equation}
\begin{multline}
 \overline{\langle\Delta n_l \, \Delta n_{l^\prime}\rangle}_{R,\textrm{dir}} =  \sum_{p}\, \overline{\langle \Delta n^R_p \Delta n_{l^\prime}\rangle}\, (P_1 + 2\,P_2 )\,,
\end{multline}
and Eqs.~(\ref{diagResCor}) and (\ref{cruzadoResCor}) are combined into 
a single result,
\begin{multline}
 \overline{\langle (\Delta n_A)^2\rangle}_{R,R} =  \sum_{p, p^\prime} \overline{\langle \Delta n^R_p\, \Delta n^R_{p^\prime}\rangle} (P_1 + 2\,P_2)(P^\prime_1 + 2\,P^\prime_2) + \\ 
+ \sum_p \overline{\langle n^R_ p \rangle} \big[P_1 + 4\,P_2 - (P_1 + 2\,P_2)^2\big] \,.
\end{multline}

A generalization of the results above to decays into more than two particles should be possible 
using similar arguments. 
A branching ratio $b_r$ can be easily introduced by making $P_l \rightarrow b_r \, P_l$.

\section{Results}
\label{secResult}

Calculations within the above scheme can be directly implemented in a computer code, which  
enables us to calculate the variances and covariances of different particle multiplicities. 

Our aim here is neither a direct comparison with experimental data  nor a test of different signatures with realistic background contributions.  
We rather examine the effect of different kinds of limitations to simple fluctuation measures and compare analytical and simulation results.  
Our background models and resonance contributions are not meant to be complete, but exemplify how different effects may be implemented and how they 
affect fluctuations. For that reason, only the decay of $\rho$ mesons into pions is considered. 

Since the mass and the coupling of pions  are better constrained, we concentrate on these particles. 
A general discussion on the limitations and advantages of our methods and results is presented in Sec.~\ref{secConclusion}.

\subsection{Signatures from pions}

\begin{figure}
  \includegraphics{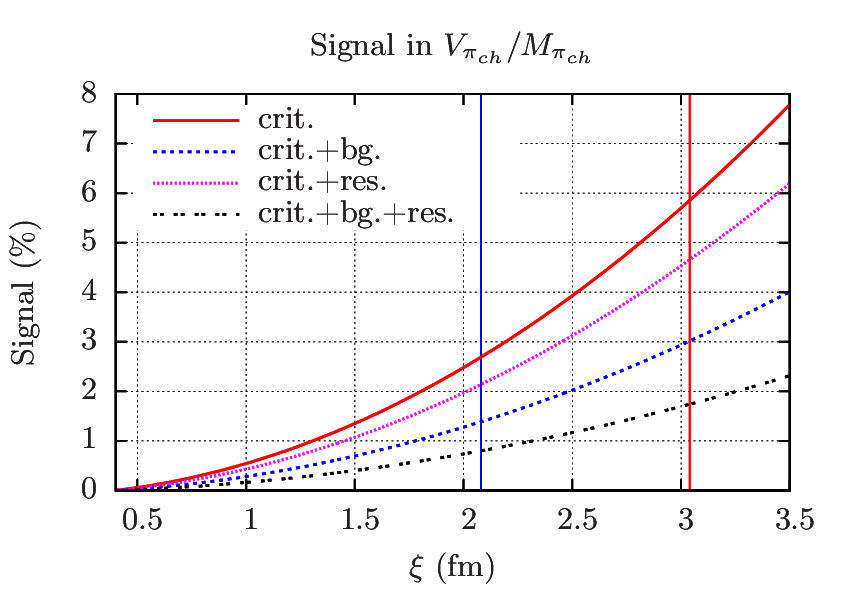}
 \caption{Signal in the variance of the charged pion multiplicity $N_{\pi^+} + N_{\pi^-}$ scaled by its average value. 
 Different curves correspond to results including (pure) critical contributions, background contributions (bg.) and/or the decay of $\rho$ resonances into two pions (res.). 
 The signal is  displayed  as a function of the correlation length at freeze-out, $\xi$,  in relation to $\xi = \xi_r = 0.4$~fm, 
 and the vertical lines correspond to the maximum possible values of $\xi$ discussed in Sec. \ref{secCritFluct}. 
 Here, the acceptance window is fixed at $|\eta|<0.5$ and $0.3< p_T < 1$~GeV. }
 \label{figtot-pion}
\end{figure}

\begin{figure}
 \includegraphics{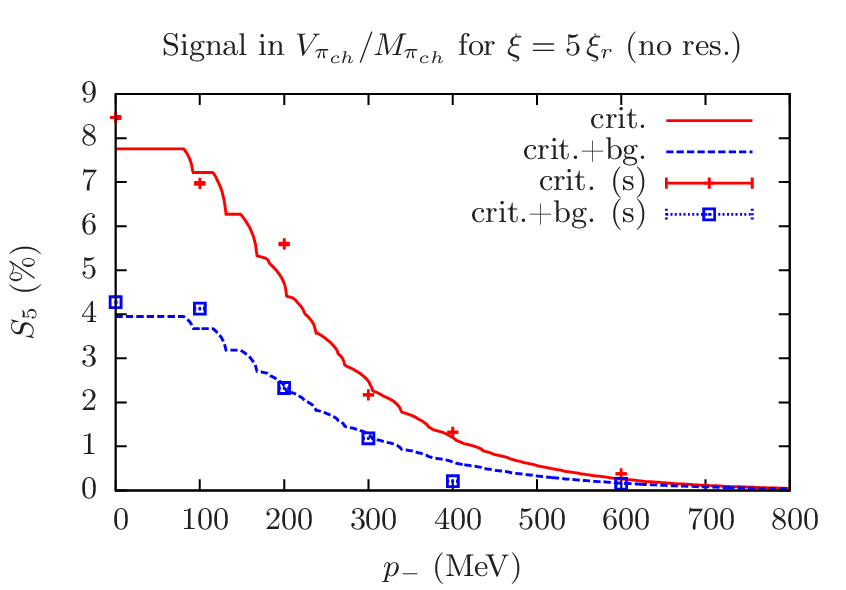}
\caption{Increase in the variance of the charged pion multiplicity $N_{\pi^+} + N_{\pi^-}$ scaled by its average value when 
the correlation length increases from $\xi_r=0.4$~fm to $\xi_c=2.0$~fm, as a function of the lower limit $p_-$ to the transverse momentum $p_T$, 
both with and without background contributions (bg.). Curves represent analytic results while points show results from numeric simulations. 
The acceptance window is given by $|\eta|<0.5$ and $p_-< p_T < 1$~GeV and contributions from resonance decay are not contemplated.}
 \label{figptmin_comp}
\end{figure}

\begin{figure}
  \includegraphics{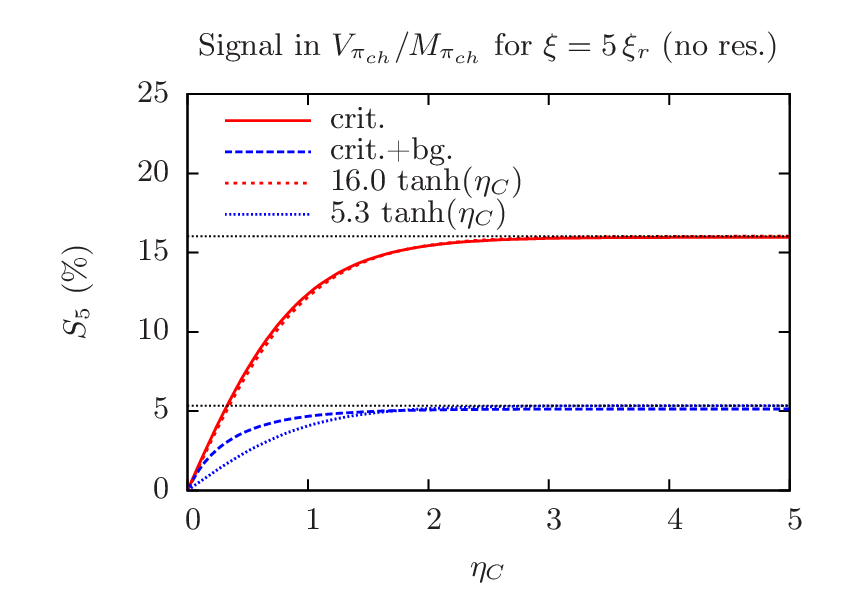}
\caption{Increase in the variance of the charged pion multiplicity $N_{\pi^+} + N_{\pi^-}$ scaled by its average value when 
the correlation length increases from $\xi_r=0.4$~fm to $\xi_c=2.0$~fm, as a function of the pseudorapidity cut $\eta_C$, 
both with and without background contributions (bg.), for $0<p_T<2$~GeV. 
Fits of the form $a \tanh (\eta_C)$ to the results are also shown.  
Contributions from resonance decay are not contemplated. }
 \label{figAcc-eta-pion}
\end{figure}

\begin{figure}
 \includegraphics{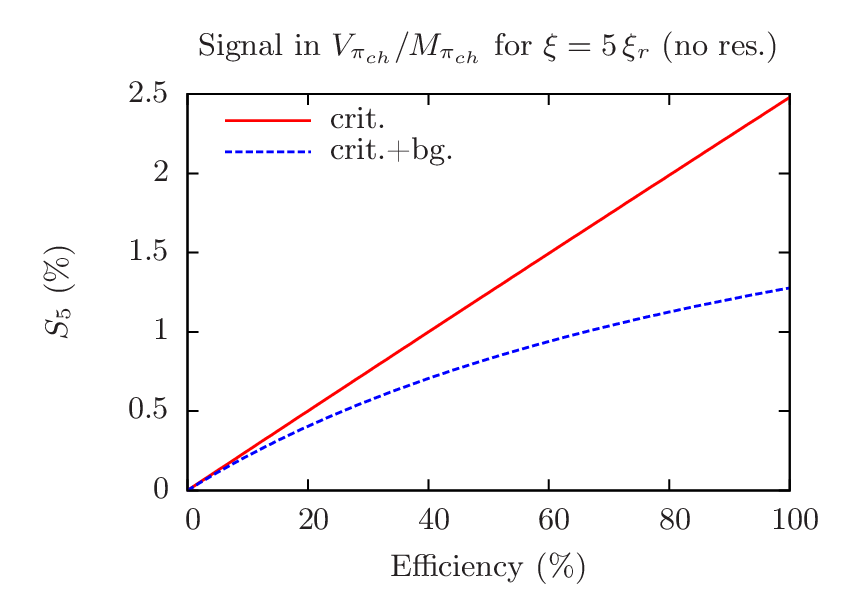}
\caption{Increase in the variance of the charged pion multiplicity $N_{\pi^+} + N_{\pi^-}$ scaled by its average value when 
the correlation length increases from $\xi_r=0.4$~fm to $\xi_c=2.0$~fm, as a function of a constant detector efficiency.  
Contributions from resonance decay are not contemplated. Results including background fluctuations (bg.) are also shown. 
The acceptance window is fixed at $|\eta|<0.5$ and $0.3< p_T < 1$~GeV and contributions from resonance decay are not contemplated.}
\label{figEff-plot}
\end{figure}

\begin{figure}
 \includegraphics{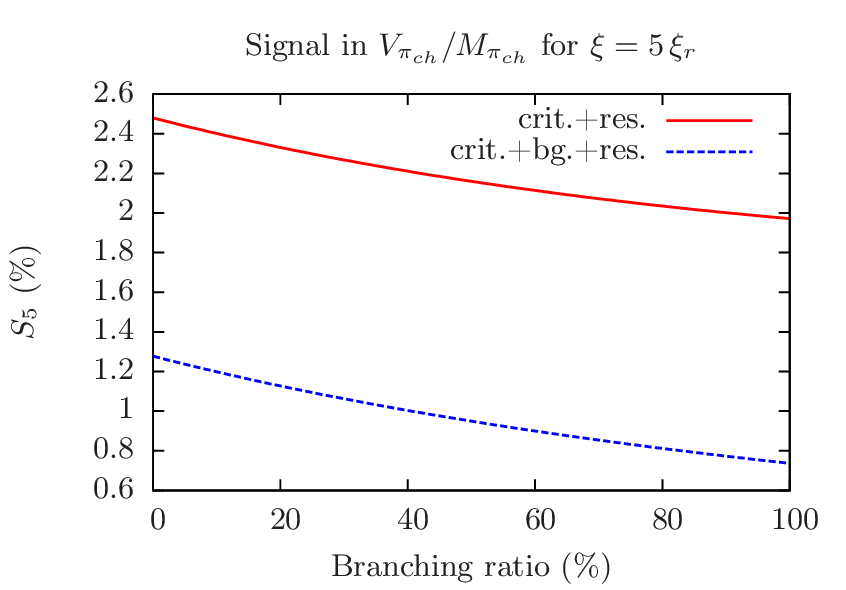}
\caption{Increase in the variance of the charged pion multiplicity $N_{\pi^+} + N_{\pi^-}$ scaled by its average value when 
the correlation length increases from $\xi_r=0.4$ fm to $\xi_c=2.0$~fm, as a function of a constant branching ratio 
for the decay of $\rho$ mesons into two pions. Results including background fluctuations (bg.) are also shown. 
The acceptance window is fixed at $|\eta|<0.5$ and $0.3< p_T < 1$~GeV and contributions from resonance decay are not contemplated.}
\label{figB-ratio}
\end{figure}

We now turn to signatures coming exclusively from charged pions and examine their dependence on the chiral correlation length $\xi$. 
As in Ref.~\cite{Hippert2016}, we use $T = 130$~MeV for the temperature of the system and $R_p = 6.8$~fm for its average radius. 
 We choose to display results for the increase in the ratio between the variance and the average of the multiplicity of charged pions, relative to its value at $\xi_r$, as a function of $\xi$:
 \begin{equation}
  \textrm{Signal} (\xi) := \dfrac{V_{\pi_{ch}}(\xi)/M_{\pi_{ch}}(\xi)}{V_{\pi_{ch}}(\xi_r)/M_{\pi_{ch}}(\xi_r)}-1\,,
\label{def-signal} 
\end{equation}
 where $M_{\pi_{ch}} = \overline{\langle N_{\pi_{ch}} \rangle}$ and $V_{\pi_{ch}} = \overline{\langle (\Delta N_{\pi_{ch}})^2 \rangle}$ are the mean and variance of the charged pion multiplicity and, 
unlike in Ref.~\cite{Hippert2016}, we use $\xi_r = 0.4$~fm $\sim 1/m_\sigma^{\textrm vac}$ as a baseline value for the correlation length. 
Figure~\ref{figtot-pion} displays results for the effects of critical and background contributions as well as the decay of $\rho$ mesons into two pions. 
These results suggest that the chosen signature would hardly reach $\sim 1\%$. 
Figures~\ref{figptmin_comp}, \ref{figAcc-eta-pion}, \ref{figEff-plot} and \ref{figB-ratio} show results for the signal at $\xi= 5 \, \xi_r = 2.0$~fm, $S_5$,  
as  a function of the lower transverse momentum cut $p_-$, the pseudorapidity cut $\eta_C$, a constant detector efficiency and the branching ratio for the decay of $\rho$ mesons into two pions, respectively.

Figure~\ref{figptmin_comp} also presents simulation results, showing good agreement with analytical ones for different transverse momentum cuts. 
This validates the approximation, made in Sec.~\ref{secKC}, that fluctuations of $F(p)$ are negligible. 
The decrease of $S_5$ with an increasing $p_-$ is due to a larger concentration of pions at low transverse momentum, which 
also makes the value of the superior transverse momentum cut $p_+$ less relevant.  

Attention should be drawn to the fact that the signature in Eq.~(\ref{def-signal}) is also affected by particle number fluctuations which are inherent to a grand-canonical ensemble. 
These fluctuations alone yield a nearly Poissonian contribution, $V_{\pi_{ch}} / M_{\pi_{ch}}\big|_{GC} \approx 1$, which dominates the variance of the particle multiplicity. 
Since our calculations only have meaning within a grand-canonical ensemble, this contribution is present in all of our results, even the ones for ``pure'' critical contributions. 
This is partly the reason why our signals grow more slowly in $\xi$ than one would naively expect. 
While subtracting a Poissonian contribution from our signature would result in much stronger relative signals, 
such signals would correspond to much less impressive values in absolute numbers, signaling to small increases in the particle number fluctuations. 
On the other hand, calculating the signal with respect to the full background contributions, including the ones which are intrinsic to a grand-canonical ensemble, 
has the advantage of providing a direct comparison to a number we know is of order $\sim {\cal O}(1)$. 
The signals obtained after subtracting a Poissonian background are related to our results by an approximately constant ratio, which is different for each case in Fig.~\ref{figtot-pion}. 
Without the additional spurious fluctuations, this ratio is of roughly $\sim 80$, but it drops to $\sim 2.2$ in their presence. 
If the decay of $\rho$ mesons is included, these values drop to $\sim 62$ and $\sim 1.7$, respectively. 
This shows that, despite growing more slowly with $\xi$, our naive signature is a lot more robust to other background contributions.

The behavior for different pseudorapidity cuts in Fig.~\ref{figAcc-eta-pion} can be understood by arguments similar to the ones in Ref.~\cite{Ling2016}, although, there, expansion 
effects are considered and background contributions are not contemplated. 
Because of the isotropy assumption, as a larger rapidity window is used, a larger number of pairs of momentum modes, 
roughly proportional to the square of the window in the cosine of the polar angle $[\Delta(\cos \theta)]^2$, 
is correlated by critical contributions. Since the results are normalized by the average multiplicity of charged pions ($\propto \Delta(\cos \theta)$), 
they scale roughly as $\propto \Delta(\cos \theta)\propto F(p) \propto \tanh \eta_C$  if $p_T$ is unrestricted, 
as can also be seen from Eq.~(\ref{varNKC}) and is confirmed by the very successful $a \tanh(\eta_C)$ fit to the results. 
Since the expansion of the system was not considered, the interpretation of these results is purely geometric.  

The dependence of $S_5$ on the detector efficiency, shown in Fig.~\ref{figEff-plot}, can be modeled by assuming a detection probability 
proportional to the efficiency $e$, so that $V_{\pi_{ch}}/M_{\pi_{ch}} = \lambda(\xi) \, e + 1$ and $S_5$ takes the form 
\begin{equation}
 S_5 = \dfrac{\kappa \, e}{\gamma \, e + 1}\,,
\end{equation}
which is in excellent agreement with results.\footnote{Fits to the results yield  $\kappa = (2.5107 \pm 0.0000003)\,10^{-2},\;\gamma =(1.25693 \pm 0.00001)\,10^{-4}$ 
without background contributions and $\kappa = (2.36801 \pm  0.0000004)\,10^{-2}, \; \gamma =(8.54053 \pm 0.000004)\,10^{-3}$ when taking them into account.}

As for the dependence of the results on the branching ratio of the decay of a $\rho$ meson into two pions, shown in Fig.~\ref{figB-ratio}, 
it can be roughly described by assuming a Poissonian distribution on the number of decay products, with a mean proportional to the 
branching ratio $b_r$.  Combining contributions from decay products with critical ones while assuming statistical independence between the two and 
neglecting the dependence of $M_{\pi_{ch}}$ in $\xi$ yields 
\begin{equation}
 S_5 = \dfrac{\chi}{\zeta\, b_r + 1}\,,
\end{equation}
which can fit the results quite well.\footnote{Fits to the results yield $\chi = 2.45388 \pm 0.002523,\;\zeta =(2.59846 \pm 0.02242)\,10^{-3}$ without 
background contributions and $\chi = 1.28827 \pm 0.00122, \; \zeta =(7.27304 \pm 0.02838)\, 10^{-3}$ when taking them into account.}

The signal in Fig.~\ref{figtot-pion} scales quadratically with the correlation length, with a proportionality factor which depends mainly on
the square of the coupling $G$ in Eq.~(\ref{eqLagInt}), as well as our models for spurious contributions and the chosen acceptance window. 
While the estimate of $G = 300$~MeV near the critical point carries large uncertainties related to medium effects
and the properties of the $\sigma$ meson, it is clear that it should be much smaller than in vacuum, by a factor of $\sim 1/6$.
Since this estimate considers a vacuum sigma mass of $600$~MeV, we believe it to be optimistic~\cite{Stephanov:1999zu}.   
Regarding spurious contributions, while our models most likely overestimate the role of temperature fluctuations, they
largely neglect the complex nature of heavy-ion collision experiments. 
Among others, the effect of indirect particles is underestimated, since we have only considered the decay of $\rho$ mesons. 
Finally, although we rely on the the simplified interaction Lagrangian of Eq.~(\ref{eqLagInt}), we believe it captures most
of the physics, at least for second-order fluctuations. Further caveats to our results are discussed in Sec.~\ref{secConclusion}. 

\section{Concluding remarks}
\label{secConclusion}

In this paper, we have outlined a simple framework for calculating multiplicity fluctuations in the neighborhood of the QCD critical point. 
As we have shown, it allows for the introduction of general long-range fluctuations, acceptance cuts and resonance decay in a systematic way. 
Acceptance window effects were treated probabilistically, giving rise to extra contributions to fluctuations, and 
 applied to resonance decay. 
Finite-size effects were also included, which could be important at low collision energies. 
Moreover, there is no reason one should not extrapolate results to the continuum limit.

While we have calculated simple signatures from Gaussian fluctuations of charged pions, %as far as we can see, 
the same treatment might as well be applied to the calculation of more interesting signatures, such as 
higher moments of the net-proton distribution. 
Nevertheless, the inclusion  of other particles, as well as an extension to non-Gaussian fluctuations, requires 
the inclusion of poorly known parameters, such as the proton mass $m_p$ and the proton coupling constant $g$ near the critical point, 
and the parameters required for an appropriate description of non-Gaussian fluctuations~\cite{Stephanov:2008qz},  
a problem which may be partially solved by somehow extracting information about fluctuations in equilibrium from a provided equation of state.

Our approach presents some important limitations. It includes no dynamics, except for the estimation of the maximum 
value of the correlation length, and no freeze-out mechanism or rescattering. 
Fluctuations of the $\sigma_0$ and other parameters are taken to be perfectly homogeneous and 
particles are assumed to obey perfect symmetry and excessively restrictive boundary conditions. 

Moreover, our background fluctuations are still unrefined, although they could be easily substituted 
for more physically and experimentally motivated models. 
Information on the statistical distribution of thermodynamic parameters may be extracted from the event-by-event distribution of 
the mean transverse momentum of particles, specially for temperature fluctuations.  
In this case, different particle species, with different masses and charges, might be used to separate effects from temperature, chemical potential and volume 
fluctuations.  
In the case of volume fluctuations, a more complete model could be achieved  by assigning a probability distribution for the proportionality factor $C$ 
in Sec.~\ref{secBGf}. 

A relevant restriction is that we assume isotropy and global equilibrium,  neglecting the effects of flow.  
This means that our results for the acceptance window dependence should not be taken at face value, especially at large pseudorapidities
(compare with~\cite{Ling2016}) and, even so, their validity is limited to central collisions. 
Although these effects were not considered, they can be roughly included by using a prescribed velocity profile 
and writing 
\begin{equation}
  F(p) = \displaystyle\int \dfrac{d^3 x}{V} \int_{\tilde\Omega_{\textrm{acc}}(p,{\bf x})}\dfrac{d\Omega}{4\,\pi}\,, 
\end{equation}
where $\tilde\Omega_{\textrm{acc}}(p,{\bf x})$ is the solid angle coverage of the acceptance window when boosted 
with the velocity of the fluid element at $\bf x$. 
This would, however, provide only a  crude solution.

The methods we have employed seem to be very adaptable, enabling several effects 
to be considered at once in reasonably simple analytical calculations and simulations which are relatively 
cheap in computer power. 
These simulations produce integer numbers of particles without introducing artificial Poissonian noise 
and might be used to feed some kind of rescattering algorithm.

Another nice feature of our simulations is that they can be adapted 
to include global conservation laws (see, for instance Refs. \cite{Bzdak2013,Begun2006,Begun2004}) by restricting them to configurations with 
fixed values for conserved quantities, discarding samples which do not follow this criterion. 
However, this would raise the computing time by increasing both the number 
of necessary samples and the number of considered momentum modes.

\acknowledgments

The authors would like to thank V. Koch, M. Lisa, L. Moriconi, D. Parganlija, P. Sorensen, M. Tannenbaum and G. Torrieri for elucidative conversations and constructive criticism, 
and the organizers and participants of the INT Program “Exploring the QCD Phase Diagram through Energy Scans" for two weeks of very inspiring discussions and presentations. We 
thank the Institute for Nuclear Theory at the University of Washington for its hospitality and the Department of Energy for partial support during the completion of this work. 
We also acknowledge CNPq and FAPERJ for their financial support.

\bibliography{paper2Bib}

\begin{thebibliography}{47}%
\makeatletter
\providecommand \@ifxundefined [1]{%
 \@ifx{#1\undefined}
}%
\providecommand \@ifnum [1]{%
 \ifnum #1\expandafter \@firstoftwo
 \else \expandafter \@secondoftwo
 \fi
}%
\providecommand \@ifx [1]{%
 \ifx #1\expandafter \@firstoftwo
 \else \expandafter \@secondoftwo
 \fi
}%
\providecommand \natexlab [1]{#1}%
\providecommand \enquote  [1]{``#1''}%
\providecommand \bibnamefont  [1]{#1}%
\providecommand \bibfnamefont [1]{#1}%
\providecommand \citenamefont [1]{#1}%
\providecommand \href@noop [0]{\@secondoftwo}%
\providecommand \href [0]{\begingroup \@sanitize@url \@href}%
\providecommand \@href[1]{\@@startlink{#1}\@@href}%
\providecommand \@@href[1]{\endgroup#1\@@endlink}%
\providecommand \@sanitize@url [0]{\catcode `\\12\catcode `\$12\catcode
  `\&12\catcode `\#12\catcode `\^12\catcode `\_12\catcode `\%12\relax}%
\providecommand \@@startlink[1]{}%
\providecommand \@@endlink[0]{}%
\providecommand \url  [0]{\begingroup\@sanitize@url \@url }%
\providecommand \@url [1]{\endgroup\@href {#1}{\urlprefix }}%
\providecommand \urlprefix  [0]{URL }%
\providecommand \Eprint [0]{\href }%
\providecommand \doibase [0]{http://dx.doi.org/}%
\providecommand \selectlanguage [0]{\@gobble}%
\providecommand \bibinfo  [0]{\@secondoftwo}%
\providecommand \bibfield  [0]{\@secondoftwo}%
\providecommand \translation [1]{[#1]}%
\providecommand \BibitemOpen [0]{}%
\providecommand \bibitemStop [0]{}%
\providecommand \bibitemNoStop [0]{.\EOS\space}%
\providecommand \EOS [0]{\spacefactor3000\relax}%
\providecommand \BibitemShut  [1]{\csname bibitem#1\endcsname}%
\let\auto@bib@innerbib\@empty
%</preamble>
\bibitem [{\citenamefont {Aoki}\ \emph {et~al.}(2006)\citenamefont {Aoki},
  \citenamefont {Endrodi}, \citenamefont {Fodor}, \citenamefont {Katz},\ and\
  \citenamefont {Szabo}}]{Aoki:2006we}%
  \BibitemOpen
  \bibfield  {author} {\bibinfo {author} {\bibfnamefont {Y.}~\bibnamefont
  {Aoki}}, \bibinfo {author} {\bibfnamefont {G.}~\bibnamefont {Endrodi}},
  \bibinfo {author} {\bibfnamefont {Z.}~\bibnamefont {Fodor}}, \bibinfo
  {author} {\bibfnamefont {S.~D.}\ \bibnamefont {Katz}}, \ and\ \bibinfo
  {author} {\bibfnamefont {K.~K.}\ \bibnamefont {Szabo}},\ }\bibfield  {title}
  {\enquote {\bibinfo {title} {{The Order of the quantum chromodynamics
  transition predicted by the standard model of particle physics}},}\ }\href
  {http://dx.doi.org/10.1038/nature05120} {\bibfield  {journal} {\bibinfo
  {journal} {Nature}\ }\textbf {\bibinfo {volume} {443}},\ \bibinfo {pages}
  {675--678} (\bibinfo {year} {2006})},\ \Eprint
  {http://arxiv.org/abs/hep-lat/0611014} {arXiv:hep-lat/0611014 [hep-lat]}
  \BibitemShut {NoStop}%
%%CITATION = HEP-LAT/0611014;%%
\bibitem [{\citenamefont {Borsanyi}\ \emph {et~al.}(2010)\citenamefont
  {Borsanyi}, \citenamefont {Endrodi}, \citenamefont {Fodor}, \citenamefont
  {Jakovac}, \citenamefont {Katz}, \citenamefont {Krieg}, \citenamefont
  {Ratti},\ and\ \citenamefont {Szabo}}]{Borsanyi:2010cj}%
  \BibitemOpen
  \bibfield  {author} {\bibinfo {author} {\bibfnamefont {S.}~\bibnamefont
  {Borsanyi}}, \bibinfo {author} {\bibfnamefont {G.}~\bibnamefont {Endrodi}},
  \bibinfo {author} {\bibfnamefont {Z.}~\bibnamefont {Fodor}}, \bibinfo
  {author} {\bibfnamefont {A.}~\bibnamefont {Jakovac}}, \bibinfo {author}
  {\bibfnamefont {S.~D.}\ \bibnamefont {Katz}}, \bibinfo {author}
  {\bibfnamefont {S.}~\bibnamefont {Krieg}}, \bibinfo {author} {\bibfnamefont
  {C.}~\bibnamefont {Ratti}}, \ and\ \bibinfo {author} {\bibfnamefont {K.~K.}\
  \bibnamefont {Szabo}},\ }\bibfield  {title} {\enquote {\bibinfo {title} {{The
  QCD equation of state with dynamical quarks}},}\ }\href
  {http://dx.doi.org/10.1007/JHEP11(2010)077} {\bibfield  {journal} {\bibinfo
  {journal} {JHEP}\ }\textbf {\bibinfo {volume} {11}},\ \bibinfo {pages} {077}
  (\bibinfo {year} {2010})},\ \Eprint {http://arxiv.org/abs/1007.2580}
  {arXiv:1007.2580 [hep-lat]} \BibitemShut {NoStop}%
%%CITATION = ARXIV:1007.2580;%%
\bibitem [{\citenamefont {Stephanov}(2006)}]{Stephanov:2007fk}%
  \BibitemOpen
  \bibfield  {author} {\bibinfo {author} {\bibfnamefont {M.~A.}\ \bibnamefont
  {Stephanov}},\ }\bibfield  {title} {\enquote {\bibinfo {title} {{QCD phase
  diagram: An Overview}},}\ }\href {https://pos.sissa.it/032/024/} {\bibfield
  {journal} {\bibinfo  {journal} {PoS}\ }\textbf {\bibinfo {volume}
  {LAT2006}},\ \bibinfo {pages} {024} (\bibinfo {year} {2006})},\ \Eprint
  {http://arxiv.org/abs/hep-lat/0701002} {arXiv:hep-lat/0701002 [hep-lat]}
  \BibitemShut {NoStop}%
%%CITATION = HEP-LAT/0701002;%%
\bibitem [{\citenamefont {Stephanov}(2004)}]{Stephanov:2004wx}%
  \BibitemOpen
  \bibfield  {author} {\bibinfo {author} {\bibfnamefont {M.~A.}\ \bibnamefont
  {Stephanov}},\ }\bibfield  {title} {\enquote {\bibinfo {title} {{QCD phase
  diagram and the critical point}},}\ }\href
  {http://dx.doi.org/10.1142/S0217751X05027965} {\bibfield  {journal} {\bibinfo
   {journal} {Prog.Theor.Phys.Suppl.}\ }\textbf {\bibinfo {volume} {153}},\
  \bibinfo {pages} {139--156} (\bibinfo {year} {2004})},\ \Eprint
  {http://arxiv.org/abs/hep-ph/0402115} {arXiv:hep-ph/0402115 [hep-ph]}
  \BibitemShut {NoStop}%
%%CITATION = HEP-PH/0402115;%%
\bibitem [{\citenamefont {{Rajagopal}}(1995)}]{Rajagopal:1995bc}%
  \BibitemOpen
  \bibfield  {author} {\bibinfo {author} {\bibfnamefont {K.}~\bibnamefont
  {{Rajagopal}}},\ }\enquote {\bibinfo {title} {{The Chiral Phase Transition in
  Qcd: Critical Phenomena and Long Wavelength Pion Oscillations}},}\ in\ \href
  {http://dx.doi.org/10.1142/9789812830661_0009} {\emph {\bibinfo {booktitle}
  {{Quark-Gluon Plasma 2.}}}},\ \bibinfo {editor} {edited by\ \bibinfo {editor}
  {\bibfnamefont {R.~C.}\ \bibnamefont {{Hwa}}}}\ (\bibinfo  {publisher} {World
  Scientific Publishing Co},\ \bibinfo {year} {1995})\ pp.\ \bibinfo {pages}
  {484--554}\BibitemShut {NoStop}%
\bibitem [{\citenamefont {Luo}(2016)}]{Luo:2015doi}%
  \BibitemOpen
  \bibfield  {author} {\bibinfo {author} {\bibfnamefont {X.}~\bibnamefont
  {Luo}},\ }\bibfield  {title} {\enquote {\bibinfo {title} {{Exploring the QCD
  Phase Structure with Beam Energy Scan in Heavy-ion Collisions}},}\ }\href
  {http://dx.doi.org/10.1016/j.nuclphysa.2016.03.025} {\bibfield  {journal}
  {\bibinfo  {journal} {Nucl. Phys.}\ }\textbf {\bibinfo {volume} {A956}},\
  \bibinfo {pages} {75--82} (\bibinfo {year} {2016})},\ \Eprint
  {http://arxiv.org/abs/1512.09215} {arXiv:1512.09215 [nucl-ex]} \BibitemShut
  {NoStop}%
%%CITATION = ARXIV:1512.09215;%%
\bibitem [{\citenamefont {Luo}(2015)}]{Luo:2015ewa}%
  \BibitemOpen
  \bibfield  {author} {\bibinfo {author} {\bibfnamefont {X.}~\bibnamefont
  {Luo}} (\bibinfo {collaboration} {for the STAR Collaboration}),\ }\bibfield
  {title} {\enquote {\bibinfo {title} {{Energy Dependence of Moments of
  Net-Proton and Net-Charge Multiplicity Distributions at STAR}},}\ }\href
  {https://pos.sissa.it/217/019/} {\bibfield  {journal} {\bibinfo  {journal}
  {PoS}\ }\textbf {\bibinfo {volume} {CPOD2014}},\ \bibinfo {pages} {019}
  (\bibinfo {year} {2015})},\ \Eprint {http://arxiv.org/abs/1503.02558}
  {arXiv:1503.02558 [nucl-ex]} \BibitemShut {NoStop}%
%%CITATION = ARXIV:1503.02558;%%
\bibitem [{\citenamefont {Xu}(2016)}]{Xu:2016mqs}%
  \BibitemOpen
  \bibfield  {author} {\bibinfo {author} {\bibfnamefont {J.}~\bibnamefont {Xu}}
  (\bibinfo {collaboration} {for the STAR Collaboration}),\ }\bibfield  {title}
  {\enquote {\bibinfo {title} {{Energy Dependence of Moments of Net-Proton,
  Net-Kaon, and Net-Charge Multiplicity Distributions at STAR}},}\ }\href
  {http://dx.doi.org/10.1088/1742-6596/736/1/012002} {\bibfield  {journal}
  {\bibinfo  {journal} {J. Phys. Conf. Ser.}\ }\textbf {\bibinfo {volume}
  {736}},\ \bibinfo {pages} {012002} (\bibinfo {year} {2016})},\ \Eprint
  {http://arxiv.org/abs/1611.07134} {arXiv:1611.07134 [hep-ex]} \BibitemShut
  {NoStop}%
%%CITATION = ARXIV:1611.07134;%%
\bibitem [{\citenamefont {Stephanov}\ \emph {et~al.}(1999)\citenamefont
  {Stephanov}, \citenamefont {Rajagopal},\ and\ \citenamefont
  {Shuryak}}]{Stephanov:1999zu}%
  \BibitemOpen
  \bibfield  {author} {\bibinfo {author} {\bibfnamefont {M.~A.}\ \bibnamefont
  {Stephanov}}, \bibinfo {author} {\bibfnamefont {K.}~\bibnamefont
  {Rajagopal}}, \ and\ \bibinfo {author} {\bibfnamefont {E.~V.}\ \bibnamefont
  {Shuryak}},\ }\bibfield  {title} {\enquote {\bibinfo {title} {{Event-by-event
  fluctuations in heavy ion collisions and the QCD critical point}},}\ }\href
  {http://dx.doi.org/10.1103/PhysRevD.60.114028} {\bibfield  {journal}
  {\bibinfo  {journal} {Phys. Rev. D}\ }\textbf {\bibinfo {volume} {60}},\
  \bibinfo {pages} {114028} (\bibinfo {year} {1999})},\ \Eprint
  {http://arxiv.org/abs/hep-ph/9903292} {arXiv:hep-ph/9903292 [hep-ph]}
  \BibitemShut {NoStop}%
%%CITATION = HEP-PH/9903292;%%
\bibitem [{\citenamefont {Stephanov}(2009)}]{Stephanov:2008qz}%
  \BibitemOpen
  \bibfield  {author} {\bibinfo {author} {\bibfnamefont {M.A.}\ \bibnamefont
  {Stephanov}},\ }\bibfield  {title} {\enquote {\bibinfo {title} {{Non-Gaussian
  fluctuations near the QCD critical point}},}\ }\href
  {http://dx.doi.org/10.1103/PhysRevLett.102.032301} {\bibfield  {journal}
  {\bibinfo  {journal} {Phys. Rev. Lett.}\ }\textbf {\bibinfo {volume} {102}},\
  \bibinfo {pages} {032301} (\bibinfo {year} {2009})},\ \Eprint
  {http://arxiv.org/abs/0809.3450} {arXiv:0809.3450 [hep-ph]} \BibitemShut
  {NoStop}%
%%CITATION = ARXIV:0809.3450;%%
\bibitem [{\citenamefont {Athanasiou}\ \emph {et~al.}(2010)\citenamefont
  {Athanasiou}, \citenamefont {Rajagopal},\ and\ \citenamefont
  {Stephanov}}]{Athanasiou:2010kw}%
  \BibitemOpen
  \bibfield  {author} {\bibinfo {author} {\bibfnamefont {C.}~\bibnamefont
  {Athanasiou}}, \bibinfo {author} {\bibfnamefont {K.}~\bibnamefont
  {Rajagopal}}, \ and\ \bibinfo {author} {\bibfnamefont {M.}~\bibnamefont
  {Stephanov}},\ }\bibfield  {title} {\enquote {\bibinfo {title} {{Using Higher
  Moments of Fluctuations and their Ratios in the Search for the QCD Critical
  Point}},}\ }\href {http://dx.doi.org/10.1103/PhysRevD.82.074008} {\bibfield
  {journal} {\bibinfo  {journal} {Phys. Rev. D}\ }\textbf {\bibinfo {volume}
  {82}},\ \bibinfo {pages} {074008} (\bibinfo {year} {2010})},\ \Eprint
  {http://arxiv.org/abs/1006.4636} {arXiv:1006.4636 [hep-ph]} \BibitemShut
  {NoStop}%
%%CITATION = ARXIV:1006.4636;%%
\bibitem [{\citenamefont {Berdnikov}\ and\ \citenamefont
  {Rajagopal}(2000)}]{Berdnikov:1999ph}%
  \BibitemOpen
  \bibfield  {author} {\bibinfo {author} {\bibfnamefont {B.}~\bibnamefont
  {Berdnikov}}\ and\ \bibinfo {author} {\bibfnamefont {K.}~\bibnamefont
  {Rajagopal}},\ }\bibfield  {title} {\enquote {\bibinfo {title} {{Slowing
  out-of-equilibrium near the QCD critical point}},}\ }\href
  {http://dx.doi.org/10.1103/PhysRevD.61.105017} {\bibfield  {journal}
  {\bibinfo  {journal} {Phys. Rev. D}\ }\textbf {\bibinfo {volume} {61}},\
  \bibinfo {pages} {105017} (\bibinfo {year} {2000})},\ \Eprint
  {http://arxiv.org/abs/hep-ph/9912274} {arXiv:hep-ph/9912274 [hep-ph]}
  \BibitemShut {NoStop}%
%%CITATION = HEP-PH/9912274;%%
\bibitem [{\citenamefont {Braun}\ \emph {et~al.}(2006)\citenamefont {Braun},
  \citenamefont {Klein}, \citenamefont {Pirner},\ and\ \citenamefont
  {Rezaeian}}]{Braun:2005fj}%
  \BibitemOpen
  \bibfield  {author} {\bibinfo {author} {\bibfnamefont {J.}~\bibnamefont
  {Braun}}, \bibinfo {author} {\bibfnamefont {B.}~\bibnamefont {Klein}},
  \bibinfo {author} {\bibfnamefont {H.~J.}\ \bibnamefont {Pirner}}, \ and\
  \bibinfo {author} {\bibfnamefont {A.~H.}\ \bibnamefont {Rezaeian}},\
  }\bibfield  {title} {\enquote {\bibinfo {title} {{Volume and quark mass
  dependence of the chiral phase transition}},}\ }\href
  {http://dx.doi.org/10.1103/PhysRevD.73.074010} {\bibfield  {journal}
  {\bibinfo  {journal} {Phys. Rev. D}\ }\textbf {\bibinfo {volume} {73}},\
  \bibinfo {pages} {074010} (\bibinfo {year} {2006})},\ \Eprint
  {http://arxiv.org/abs/hep-ph/0512274} {arXiv:hep-ph/0512274 [hep-ph]}
  \BibitemShut {NoStop}%
%%CITATION = HEP-PH/0512274;%%
\bibitem [{\citenamefont {Kiriyama}\ \emph {et~al.}(2006)\citenamefont
  {Kiriyama}, \citenamefont {Kodama},\ and\ \citenamefont
  {Koide}}]{Kiriyama:2006uh}%
  \BibitemOpen
  \bibfield  {author} {\bibinfo {author} {\bibfnamefont {O.}~\bibnamefont
  {Kiriyama}}, \bibinfo {author} {\bibfnamefont {T.}~\bibnamefont {Kodama}}, \
  and\ \bibinfo {author} {\bibfnamefont {T.}~\bibnamefont {Koide}},\ }\bibfield
   {title} {\enquote {\bibinfo {title} {{Finite-size effects on the QCD phase
  diagram}},}\ }\href@noop {} {\  (\bibinfo {year} {2006})},\ \Eprint
  {http://arxiv.org/abs/hep-ph/0602086} {arXiv:hep-ph/0602086 [hep-ph]}
  \BibitemShut {NoStop}%
%%CITATION = HEP-PH/0602086;%%
\bibitem [{\citenamefont {Palhares}\ \emph {et~al.}(2011)\citenamefont
  {Palhares}, \citenamefont {Fraga},\ and\ \citenamefont
  {Kodama}}]{Palhares:2009tf}%
  \BibitemOpen
  \bibfield  {author} {\bibinfo {author} {\bibfnamefont {L.~F.}\ \bibnamefont
  {Palhares}}, \bibinfo {author} {\bibfnamefont {E.~S.}\ \bibnamefont {Fraga}},
  \ and\ \bibinfo {author} {\bibfnamefont {T.}~\bibnamefont {Kodama}},\
  }\bibfield  {title} {\enquote {\bibinfo {title} {{Chiral transition in a
  finite system and possible use of finite size scaling in relativistic heavy
  ion collisions}},}\ }\href {http://dx.doi.org/10.1088/0954-3899/38/8/085101}
  {\bibfield  {journal} {\bibinfo  {journal} {J. Phys. G}\ }\textbf {\bibinfo
  {volume} {38}},\ \bibinfo {pages} {085101} (\bibinfo {year} {2011})},\
  \Eprint {http://arxiv.org/abs/0904.4830} {arXiv:0904.4830 [nucl-th]}
  \BibitemShut {NoStop}%
%%CITATION = ARXIV:0904.4830;%%
\bibitem [{\citenamefont {Fraga}\ \emph {et~al.}(2011)\citenamefont {Fraga},
  \citenamefont {Palhares},\ and\ \citenamefont {Sorensen}}]{Fraga:2011hi}%
  \BibitemOpen
  \bibfield  {author} {\bibinfo {author} {\bibfnamefont {E.~S.}\ \bibnamefont
  {Fraga}}, \bibinfo {author} {\bibfnamefont {L.~F.}\ \bibnamefont {Palhares}},
  \ and\ \bibinfo {author} {\bibfnamefont {P.}~\bibnamefont {Sorensen}},\
  }\bibfield  {title} {\enquote {\bibinfo {title} {{Finite-size scaling as a
  tool in the search for the QCD critical point in heavy ion data}},}\ }\href
  {http://dx.doi.org/10.1103/PhysRevC.84.011903} {\bibfield  {journal}
  {\bibinfo  {journal} {Phys. Rev. C}\ }\textbf {\bibinfo {volume} {84}},\
  \bibinfo {pages} {011903} (\bibinfo {year} {2011})},\ \Eprint
  {http://arxiv.org/abs/1104.3755} {arXiv:1104.3755 [hep-ph]} \BibitemShut
  {NoStop}%
%%CITATION = ARXIV:1104.3755;%%
\bibitem [{\citenamefont {Mukherjee}\ \emph {et~al.}(2015)\citenamefont
  {Mukherjee}, \citenamefont {Venugopalan},\ and\ \citenamefont
  {Yin}}]{Mukherjee2015}%
  \BibitemOpen
  \bibfield  {author} {\bibinfo {author} {\bibfnamefont {S.}~\bibnamefont
  {Mukherjee}}, \bibinfo {author} {\bibfnamefont {R.}~\bibnamefont
  {Venugopalan}}, \ and\ \bibinfo {author} {\bibfnamefont {Y.}~\bibnamefont
  {Yin}},\ }\bibfield  {title} {\enquote {\bibinfo {title} {{Real time
  evolution of non-Gaussian cumulants in the QCD critical regime}},}\ }\href
  {http://dx.doi.org/10.1103/PhysRevC.92.034912} {\bibfield  {journal}
  {\bibinfo  {journal} {Phys. Rev. C}\ }\textbf {\bibinfo {volume} {92}},\
  \bibinfo {pages} {034912} (\bibinfo {year} {2015})},\ \Eprint
  {http://arxiv.org/abs/1506.00645} {arXiv:1506.00645 [hep-ph]} \BibitemShut
  {NoStop}%
%%CITATION = ARXIV:1506.00645;%%
\bibitem [{\citenamefont {Mukherjee}\ \emph {et~al.}(2016)\citenamefont
  {Mukherjee}, \citenamefont {Venugopalan},\ and\ \citenamefont
  {Yin}}]{Mukherjee2016}%
  \BibitemOpen
  \bibfield  {author} {\bibinfo {author} {\bibfnamefont {S.}~\bibnamefont
  {Mukherjee}}, \bibinfo {author} {\bibfnamefont {R.}~\bibnamefont
  {Venugopalan}}, \ and\ \bibinfo {author} {\bibfnamefont {Y.}~\bibnamefont
  {Yin}},\ }\bibfield  {title} {\enquote {\bibinfo {title} {{Universal
  off-equilibrium scaling of critical cumulants in the QCD phase diagram}},}\
  }\href {http://dx.doi.org/10.1103/PhysRevLett.117.222301} {\bibfield
  {journal} {\bibinfo  {journal} {Phys. Rev. Lett.}\ }\textbf {\bibinfo
  {volume} {117}},\ \bibinfo {pages} {222301} (\bibinfo {year} {2016})},\
  \Eprint {http://arxiv.org/abs/1605.09341} {arXiv:1605.09341 [hep-ph]}
  \BibitemShut {NoStop}%
%%CITATION = ARXIV:1605.09341;%%
\bibitem [{\citenamefont {Hippert}\ \emph {et~al.}(2016)\citenamefont
  {Hippert}, \citenamefont {Fraga},\ and\ \citenamefont
  {Santos}}]{Hippert2016}%
  \BibitemOpen
  \bibfield  {author} {\bibinfo {author} {\bibfnamefont {Maurício}\
  \bibnamefont {Hippert}}, \bibinfo {author} {\bibfnamefont {Eduardo~S.}\
  \bibnamefont {Fraga}}, \ and\ \bibinfo {author} {\bibfnamefont {Edivaldo~M.}\
  \bibnamefont {Santos}},\ }\bibfield  {title} {\enquote {\bibinfo {title}
  {{Critical versus spurious fluctuations in the search for the QCD critical
  point}},}\ }\href {http://dx.doi.org/10.1103/PhysRevD.93.014029} {\bibfield
  {journal} {\bibinfo  {journal} {Phys. Rev. D}\ }\textbf {\bibinfo {volume}
  {93}},\ \bibinfo {pages} {014029} (\bibinfo {year} {2016})},\ \bibinfo {note}
  {[Phys. Rev.D93,014029(2016)]},\ \Eprint {http://arxiv.org/abs/1507.04764}
  {arXiv:1507.04764 [hep-ph]} \BibitemShut {NoStop}%
%%CITATION = ARXIV:1507.04764;%%
\bibitem [{\citenamefont {Gorenstein}(2015)}]{Gorenstein2015}%
  \BibitemOpen
  \bibfield  {author} {\bibinfo {author} {\bibfnamefont {M.}~\bibnamefont
  {Gorenstein}},\ }\bibfield  {title} {\enquote {\bibinfo {title} {{New
  Theoretical Results on Event-by-Event Fluctuations}},}\ }\href
  {https://pos.sissa.it/217/017/} {\bibfield  {journal} {\bibinfo  {journal}
  {PoS}\ }\textbf {\bibinfo {volume} {CPOD2014}},\ \bibinfo {pages} {017}
  (\bibinfo {year} {2015})},\ \Eprint {http://arxiv.org/abs/1505.04135}
  {arXiv:1505.04135 [nucl-th]} \BibitemShut {NoStop}%
%%CITATION = ARXIV:1505.04135;%%
\bibitem [{\citenamefont {Sahoo}\ \emph {et~al.}(2013)\citenamefont {Sahoo},
  \citenamefont {De},\ and\ \citenamefont {Nayak}}]{Sahoo2013}%
  \BibitemOpen
  \bibfield  {author} {\bibinfo {author} {\bibfnamefont {N.~R.}\ \bibnamefont
  {Sahoo}}, \bibinfo {author} {\bibfnamefont {S.}~\bibnamefont {De}}, \ and\
  \bibinfo {author} {\bibfnamefont {T.~K.}\ \bibnamefont {Nayak}},\ }\bibfield
  {title} {\enquote {\bibinfo {title} {{Baseline study for higher moments of
  net-charge distributions at energies available at the BNL Relativistic Heavy
  Ion Collider}},}\ }\href {http://dx.doi.org/10.1103/PhysRevC.87.044906}
  {\bibfield  {journal} {\bibinfo  {journal} {Phys. Rev. C}\ }\textbf {\bibinfo
  {volume} {87}},\ \bibinfo {pages} {044906} (\bibinfo {year} {2013})},\
  \Eprint {http://arxiv.org/abs/1210.7206} {arXiv:1210.7206 [nucl-ex]}
  \BibitemShut {NoStop}%
%%CITATION = ARXIV:1210.7206;%%
\bibitem [{\citenamefont {Herold}\ \emph {et~al.}(2016)\citenamefont {Herold},
  \citenamefont {Nahrgang}, \citenamefont {Yan},\ and\ \citenamefont
  {Kobdaj}}]{Herold2016}%
  \BibitemOpen
  \bibfield  {author} {\bibinfo {author} {\bibfnamefont {C.}~\bibnamefont
  {Herold}}, \bibinfo {author} {\bibfnamefont {M.}~\bibnamefont {Nahrgang}},
  \bibinfo {author} {\bibfnamefont {Y.}~\bibnamefont {Yan}}, \ and\ \bibinfo
  {author} {\bibfnamefont {C.}~\bibnamefont {Kobdaj}},\ }\bibfield  {title}
  {\enquote {\bibinfo {title} {{Dynamical net-proton fluctuations near a QCD
  critical point}},}\ }\href {http://dx.doi.org/10.1103/PhysRevC.93.021902}
  {\bibfield  {journal} {\bibinfo  {journal} {Phys. Rev. C}\ }\textbf {\bibinfo
  {volume} {93}},\ \bibinfo {pages} {021902} (\bibinfo {year} {2016})},\
  \Eprint {http://arxiv.org/abs/1601.04839} {arXiv:1601.04839 [hep-ph]}
  \BibitemShut {NoStop}%
%%CITATION = ARXIV:1601.04839;%%
\bibitem [{\citenamefont {Luo}\ \emph {et~al.}(2013)\citenamefont {Luo},
  \citenamefont {Xu}, \citenamefont {Mohanty},\ and\ \citenamefont
  {Xu}}]{Luo2013a}%
  \BibitemOpen
  \bibfield  {author} {\bibinfo {author} {\bibfnamefont {X.}~\bibnamefont
  {Luo}}, \bibinfo {author} {\bibfnamefont {Ji}~\bibnamefont {Xu}}, \bibinfo
  {author} {\bibfnamefont {B.}~\bibnamefont {Mohanty}}, \ and\ \bibinfo
  {author} {\bibfnamefont {N.}~\bibnamefont {Xu}},\ }\bibfield  {title}
  {\enquote {\bibinfo {title} {{Volume fluctuation and auto-correlation effects
  in the moment analysis of net-proton multiplicity distributions in heavy-ion
  collisions}},}\ }\href {http://dx.doi.org/10.1088/0954-3899/40/10/105104}
  {\bibfield  {journal} {\bibinfo  {journal} {J. Phys. G}\ }\textbf {\bibinfo
  {volume} {40}},\ \bibinfo {pages} {105104} (\bibinfo {year} {2013})},\
  \Eprint {http://arxiv.org/abs/1302.2332} {arXiv:1302.2332 [nucl-ex]}
  \BibitemShut {NoStop}%
%%CITATION = ARXIV:1302.2332;%%
\bibitem [{\citenamefont {Antoniou}\ \emph {et~al.}(2007)\citenamefont
  {Antoniou}, \citenamefont {Diakonos},\ and\ \citenamefont
  {Saridakis}}]{Antoniou2007}%
  \BibitemOpen
  \bibfield  {author} {\bibinfo {author} {\bibfnamefont {N.~G.}\ \bibnamefont
  {Antoniou}}, \bibinfo {author} {\bibfnamefont {F.~K.}\ \bibnamefont
  {Diakonos}}, \ and\ \bibinfo {author} {\bibfnamefont {E.~N.}\ \bibnamefont
  {Saridakis}},\ }\bibfield  {title} {\enquote {\bibinfo {title} {{Evolution of
  Critical Correlations at the QCD Phase Transition}},}\ }\href
  {http://dx.doi.org/10.1016/j.nuclphysa.2006.12.006} {\bibfield  {journal}
  {\bibinfo  {journal} {Nucl. Phys.}\ }\textbf {\bibinfo {volume} {A784}},\
  \bibinfo {pages} {536} (\bibinfo {year} {2007})},\ \Eprint
  {http://arxiv.org/abs/hep-ph/0610382} {arXiv:hep-ph/0610382 [hep-ph]}
  \BibitemShut {NoStop}%
%%CITATION = HEP-PH/0610382;%%
\bibitem [{\citenamefont {Antoniou}\ \emph {et~al.}(2008)\citenamefont
  {Antoniou}, \citenamefont {Diakonos},\ and\ \citenamefont
  {Saridakis}}]{Antoniou2008}%
  \BibitemOpen
  \bibfield  {author} {\bibinfo {author} {\bibfnamefont {N.~G.}\ \bibnamefont
  {Antoniou}}, \bibinfo {author} {\bibfnamefont {F.~K.}\ \bibnamefont
  {Diakonos}}, \ and\ \bibinfo {author} {\bibfnamefont {E.~N.}\ \bibnamefont
  {Saridakis}},\ }\bibfield  {title} {\enquote {\bibinfo {title} {{Evolutionary
  intermittency and the QCD critical point}},}\ }\href
  {http://dx.doi.org/10.1103/PhysRevC.78.024908} {\bibfield  {journal}
  {\bibinfo  {journal} {Phys. Rev. C}\ }\textbf {\bibinfo {volume} {78}},\
  \bibinfo {pages} {024908} (\bibinfo {year} {2008})},\ \Eprint
  {http://arxiv.org/abs/0709.0339} {arXiv:0709.0339 [hep-ph]} \BibitemShut
  {NoStop}%
%%CITATION = ARXIV:0709.0339;%%
\bibitem [{\citenamefont {Stephanov}(2010)}]{Stephanov2010}%
  \BibitemOpen
  \bibfield  {author} {\bibinfo {author} {\bibfnamefont {M.~A.}\ \bibnamefont
  {Stephanov}},\ }\bibfield  {title} {\enquote {\bibinfo {title} {{Evolution of
  fluctuations near QCD critical point}},}\ }\href
  {http://dx.doi.org/10.1103/PhysRevD.81.054012} {\bibfield  {journal}
  {\bibinfo  {journal} {Phys. Rev. D}\ }\textbf {\bibinfo {volume} {81}},\
  \bibinfo {pages} {054012} (\bibinfo {year} {2010})},\ \Eprint
  {http://arxiv.org/abs/0911.1772} {arXiv:0911.1772 [hep-ph]} \BibitemShut
  {NoStop}%
%%CITATION = ARXIV:0911.1772;%%
\bibitem [{\citenamefont {Bzdak}\ and\ \citenamefont {Koch}(2012)}]{Bzdak2012}%
  \BibitemOpen
  \bibfield  {author} {\bibinfo {author} {\bibfnamefont {A.}~\bibnamefont
  {Bzdak}}\ and\ \bibinfo {author} {\bibfnamefont {V.}~\bibnamefont {Koch}},\
  }\bibfield  {title} {\enquote {\bibinfo {title} {{Acceptance corrections to
  net baryon and net charge cumulants}},}\ }\href
  {http://dx.doi.org/10.1103/PhysRevC.86.044904} {\bibfield  {journal}
  {\bibinfo  {journal} {Phys. Rev. C}\ }\textbf {\bibinfo {volume} {86}},\
  \bibinfo {pages} {044904} (\bibinfo {year} {2012})},\ \Eprint
  {http://arxiv.org/abs/1206.4286} {arXiv:1206.4286 [nucl-th]} \BibitemShut
  {NoStop}%
%%CITATION = ARXIV:1206.4286;%%
\bibitem [{\citenamefont {Ling}\ and\ \citenamefont
  {Stephanov}(2016)}]{Ling2016}%
  \BibitemOpen
  \bibfield  {author} {\bibinfo {author} {\bibfnamefont {B.}~\bibnamefont
  {Ling}}\ and\ \bibinfo {author} {\bibfnamefont {M.~A.}\ \bibnamefont
  {Stephanov}},\ }\bibfield  {title} {\enquote {\bibinfo {title} {{Acceptance
  dependence of fluctuation measures near the QCD critical point}},}\ }\href
  {http://dx.doi.org/10.1103/PhysRevC.93.034915} {\bibfield  {journal}
  {\bibinfo  {journal} {Phys. Rev. C}\ }\textbf {\bibinfo {volume} {93}},\
  \bibinfo {pages} {034915} (\bibinfo {year} {2016})},\ \Eprint
  {http://arxiv.org/abs/1512.09125} {arXiv:1512.09125 [nucl-th]} \BibitemShut
  {NoStop}%
%%CITATION = ARXIV:1512.09125;%%
\bibitem [{\citenamefont {Gell-Mann}\ and\ \citenamefont
  {Levy}(1960)}]{GellMann:1960np}%
  \BibitemOpen
  \bibfield  {author} {\bibinfo {author} {\bibfnamefont {M.}~\bibnamefont
  {Gell-Mann}}\ and\ \bibinfo {author} {\bibfnamefont {M.}~\bibnamefont
  {Levy}},\ }\bibfield  {title} {\enquote {\bibinfo {title} {{The axial vector
  current in beta decay}},}\ }\href {http://dx.doi.org/10.1007/BF02859738}
  {\bibfield  {journal} {\bibinfo  {journal} {Nuovo Cim.}\ }\textbf {\bibinfo
  {volume} {16}},\ \bibinfo {pages} {705} (\bibinfo {year} {1960})}\BibitemShut
  {NoStop}%
%%CITATION = NUCIA,16,705;%%
\bibitem [{\citenamefont {Zinn-Justin}()}]{ZinnJustin:1998ci}%
  \BibitemOpen
  \bibfield  {author} {\bibinfo {author} {\bibfnamefont {J.}~\bibnamefont
  {Zinn-Justin}},\ }\bibfield  {title} {\enquote {\bibinfo {title}
  {{Determination of critical exponents and equation of state by field theory
  method}},}\ }\bibfield  {booktitle} {\emph {\bibinfo {booktitle} {Proceedings
  of the 6th International Conference, PI'98 - "6th International Conference on
  Path integrals from PeV to TeV: 50 years after Feynman's paper.", Florence,
  Italy, August 25-29, 1998}},\ }\href@noop {} {\ }\Eprint
  {http://arxiv.org/abs/hep-th/9810193} {arXiv:hep-th/9810193 [hep-th]}
  \BibitemShut {NoStop}%
%%CITATION = HEP-TH/9810193;%%
\bibitem [{\citenamefont {Guida}\ and\ \citenamefont
  {Zinn-Justin}(1997)}]{Guida:1996ep}%
  \BibitemOpen
  \bibfield  {author} {\bibinfo {author} {\bibfnamefont {R.}~\bibnamefont
  {Guida}}\ and\ \bibinfo {author} {\bibfnamefont {J.}~\bibnamefont
  {Zinn-Justin}},\ }\bibfield  {title} {\enquote {\bibinfo {title} {{3-D Ising
  model: The Scaling equation of state}},}\ }\href
  {http://dx.doi.org/10.1016/S0550-3213(96)00704-3} {\bibfield  {journal}
  {\bibinfo  {journal} {Nucl. Phys.}\ }\textbf {\bibinfo {volume} {B489}},\
  \bibinfo {pages} {626--652} (\bibinfo {year} {1997})},\ \Eprint
  {http://arxiv.org/abs/hep-th/9610223} {arXiv:hep-th/9610223 [hep-th]}
  \BibitemShut {NoStop}%
%%CITATION = HEP-TH/9610223;%%
\bibitem [{\citenamefont {Hohenberg}\ and\ \citenamefont
  {Halperin}(1977)}]{Hohenberg:1977ym}%
  \BibitemOpen
  \bibfield  {author} {\bibinfo {author} {\bibfnamefont {P.~C.}\ \bibnamefont
  {Hohenberg}}\ and\ \bibinfo {author} {\bibfnamefont {B.~I.}\ \bibnamefont
  {Halperin}},\ }\bibfield  {title} {\enquote {\bibinfo {title} {{Theory of
  Dynamic Critical Phenomena}},}\ }\href
  {http://dx.doi.org/10.1103/RevModPhys.49.435} {\bibfield  {journal} {\bibinfo
   {journal} {Rev. Mod. Phys.}\ }\textbf {\bibinfo {volume} {49}},\ \bibinfo
  {pages} {435--479} (\bibinfo {year} {1977})}\BibitemShut {NoStop}%
%%CITATION = RMPHA,49,435;%%
\bibitem [{\citenamefont {Luo}\ and\ \citenamefont {Xu}(2017)}]{Luo:2017faz}%
  \BibitemOpen
  \bibfield  {author} {\bibinfo {author} {\bibfnamefont {X.}~\bibnamefont
  {Luo}}\ and\ \bibinfo {author} {\bibfnamefont {N.}~\bibnamefont {Xu}},\
  }\bibfield  {title} {\enquote {\bibinfo {title} {{Search for the QCD Critical
  Point with Fluctuations of Conserved Quantities in Relativistic Heavy-Ion
  Collisions at RHIC : An Overview}},}\ }\href
  {http://dx.doi.org/10.1007/s41365-017-0257-0} {\bibfield  {journal} {\bibinfo
   {journal} {Nucl. Sci. Tech.}\ }\textbf {\bibinfo {volume} {28}},\ \bibinfo
  {pages} {112} (\bibinfo {year} {2017})},\ \Eprint
  {http://arxiv.org/abs/1701.02105} {arXiv:1701.02105 [nucl-ex]} \BibitemShut
  {NoStop}%
%%CITATION = ARXIV:1701.02105;%%
\bibitem [{\citenamefont {Vries}\ \emph {et~al.}(1987)\citenamefont {Vries},
  \citenamefont {Jager},\ and\ \citenamefont {Vries}}]{DeVries1987495}%
  \BibitemOpen
  \bibfield  {author} {\bibinfo {author} {\bibfnamefont {H.~De}\ \bibnamefont
  {Vries}}, \bibinfo {author} {\bibfnamefont {C.~W.~De}\ \bibnamefont {Jager}},
  \ and\ \bibinfo {author} {\bibfnamefont {C.~De}\ \bibnamefont {Vries}},\
  }\bibfield  {title} {\enquote {\bibinfo {title} {Nuclear
  charge-density-distribution parameters from elastic electron scattering},}\
  }\href {http://dx.doi.org/10.1016/0092-640X(87)90013-1} {\bibfield  {journal}
  {\bibinfo  {journal} {At. Data Nucl. Data Tables}\ }\textbf {\bibinfo
  {volume} {36}},\ \bibinfo {pages} {495} (\bibinfo {year} {1987})}\BibitemShut
  {NoStop}%
\bibitem [{\citenamefont {Jager}\ \emph {et~al.}(1974)\citenamefont {Jager},
  \citenamefont {Vries},\ and\ \citenamefont {Vries}}]{DeJager1974479}%
  \BibitemOpen
  \bibfield  {author} {\bibinfo {author} {\bibfnamefont {C.W.~De}\ \bibnamefont
  {Jager}}, \bibinfo {author} {\bibfnamefont {H.~De}\ \bibnamefont {Vries}}, \
  and\ \bibinfo {author} {\bibfnamefont {C.~De}\ \bibnamefont {Vries}},\
  }\bibfield  {title} {\enquote {\bibinfo {title} {Nuclear charge- and
  magnetization-density-distribution parameters from elastic electron
  scattering},}\ }\href {http://dx.doi.org/10.1016/S0092-640X(74)80002-1}
  {\bibfield  {journal} {\bibinfo  {journal} {At. Data Nucl. Data Tables}\
  }\textbf {\bibinfo {volume} {14}},\ \bibinfo {pages} {479} (\bibinfo {year}
  {1974})}\BibitemShut {NoStop}%
\bibitem [{\citenamefont {Skokov}\ \emph {et~al.}(2013)\citenamefont {Skokov},
  \citenamefont {Friman},\ and\ \citenamefont {Redlich}}]{Skokov2012}%
  \BibitemOpen
  \bibfield  {author} {\bibinfo {author} {\bibfnamefont {V.}~\bibnamefont
  {Skokov}}, \bibinfo {author} {\bibfnamefont {B.}~\bibnamefont {Friman}}, \
  and\ \bibinfo {author} {\bibfnamefont {K.}~\bibnamefont {Redlich}},\
  }\bibfield  {title} {\enquote {\bibinfo {title} {{Volume Fluctuations and
  Higher Order Cumulants of the Net Baryon Number}},}\ }\href
  {http://dx.doi.org/10.1103/PhysRevC.88.034911} {\bibfield  {journal}
  {\bibinfo  {journal} {Phys. Rev. C}\ }\textbf {\bibinfo {volume} {88}},\
  \bibinfo {pages} {034911} (\bibinfo {year} {2013})},\ \Eprint
  {http://arxiv.org/abs/1205.4756} {arXiv:1205.4756 [hep-ph]} \BibitemShut
  {NoStop}%
%%CITATION = ARXIV:1205.4756;%%
\bibitem [{\citenamefont {Gorenstein}\ and\ \citenamefont
  {Gazdzicki}(2011)}]{Gorenstein2011}%
  \BibitemOpen
  \bibfield  {author} {\bibinfo {author} {\bibfnamefont {M.~I.}\ \bibnamefont
  {Gorenstein}}\ and\ \bibinfo {author} {\bibfnamefont {M.}~\bibnamefont
  {Gazdzicki}},\ }\bibfield  {title} {\enquote {\bibinfo {title} {{Strongly
  Intensive Quantities}},}\ }\href
  {http://dx.doi.org/10.1103/PhysRevC.84.014904} {\bibfield  {journal}
  {\bibinfo  {journal} {Phys. Rev. C}\ }\textbf {\bibinfo {volume} {84}},\
  \bibinfo {pages} {014904} (\bibinfo {year} {2011})},\ \Eprint
  {http://arxiv.org/abs/1101.4865} {arXiv:1101.4865 [nucl-th]} \BibitemShut
  {NoStop}%
%%CITATION = ARXIV:1101.4865;%%
\bibitem [{\citenamefont {Karsch}\ \emph {et~al.}(2016)\citenamefont {Karsch},
  \citenamefont {Morita},\ and\ \citenamefont {Redlich}}]{Karsch2015}%
  \BibitemOpen
  \bibfield  {author} {\bibinfo {author} {\bibfnamefont {F.}~\bibnamefont
  {Karsch}}, \bibinfo {author} {\bibfnamefont {K.}~\bibnamefont {Morita}}, \
  and\ \bibinfo {author} {\bibfnamefont {K.}~\bibnamefont {Redlich}},\
  }\bibfield  {title} {\enquote {\bibinfo {title} {{Effects of kinematic cuts
  on net-electric charge fluctuations}},}\ }\href
  {http://dx.doi.org/10.1103/PhysRevC.93.034907} {\bibfield  {journal}
  {\bibinfo  {journal} {Phys. Rev. C}\ }\textbf {\bibinfo {volume} {93}},\
  \bibinfo {pages} {034907} (\bibinfo {year} {2016})},\ \Eprint
  {http://arxiv.org/abs/1508.02614} {arXiv:1508.02614 [hep-ph]} \BibitemShut
  {NoStop}%
%%CITATION = ARXIV:1508.02614;%%
\bibitem [{\citenamefont {Garg}\ \emph {et~al.}(2013)\citenamefont {Garg},
  \citenamefont {Mishra}, \citenamefont {Netrakanti}, \citenamefont {Mohanty},
  \citenamefont {Mohanty}, \citenamefont {Singh},\ and\ \citenamefont
  {Xu}}]{Garg2013}%
  \BibitemOpen
  \bibfield  {author} {\bibinfo {author} {\bibfnamefont {P.}~\bibnamefont
  {Garg}}, \bibinfo {author} {\bibfnamefont {D.~K.}\ \bibnamefont {Mishra}},
  \bibinfo {author} {\bibfnamefont {P.~K.}\ \bibnamefont {Netrakanti}},
  \bibinfo {author} {\bibfnamefont {B.}~\bibnamefont {Mohanty}}, \bibinfo
  {author} {\bibfnamefont {A.~K.}\ \bibnamefont {Mohanty}}, \bibinfo {author}
  {\bibfnamefont {B.~K.}\ \bibnamefont {Singh}}, \ and\ \bibinfo {author}
  {\bibfnamefont {N.}~\bibnamefont {Xu}},\ }\bibfield  {title} {\enquote
  {\bibinfo {title} {{Conserved number fluctuations in a hadron resonance gas
  model}},}\ }\href {http://dx.doi.org/10.1016/j.physletb.2013.09.019}
  {\bibfield  {journal} {\bibinfo  {journal} {Phys. Lett. B}\ }\textbf
  {\bibinfo {volume} {726}},\ \bibinfo {pages} {691--696} (\bibinfo {year}
  {2013})},\ \Eprint {http://arxiv.org/abs/1304.7133} {arXiv:1304.7133
  [nucl-ex]} \BibitemShut {NoStop}%
%%CITATION = ARXIV:1304.7133;%%
\bibitem [{\citenamefont {Pruneau}\ \emph {et~al.}(2002)\citenamefont
  {Pruneau}, \citenamefont {Gavin},\ and\ \citenamefont
  {Voloshin}}]{PhysRevC.66.044904}%
  \BibitemOpen
  \bibfield  {author} {\bibinfo {author} {\bibfnamefont {C.}~\bibnamefont
  {Pruneau}}, \bibinfo {author} {\bibfnamefont {S.}~\bibnamefont {Gavin}}, \
  and\ \bibinfo {author} {\bibfnamefont {S.}~\bibnamefont {Voloshin}},\
  }\bibfield  {title} {\enquote {\bibinfo {title} {{Methods for the study of
  particle production fluctuations}},}\ }\href
  {http://dx.doi.org/10.1103/PhysRevC.66.044904} {\bibfield  {journal}
  {\bibinfo  {journal} {Phys. Rev. C}\ }\textbf {\bibinfo {volume} {66}},\
  \bibinfo {pages} {044904} (\bibinfo {year} {2002})},\ \Eprint
  {http://arxiv.org/abs/nucl-ex/0204011} {arXiv:nucl-ex/0204011 [nucl-ex]}
  \BibitemShut {NoStop}%
%%CITATION = NUCL-EX/0204011;%%
\bibitem [{\citenamefont {Bluhm}\ \emph {et~al.}(2017)\citenamefont {Bluhm},
  \citenamefont {Nahrgang}, \citenamefont {Bass},\ and\ \citenamefont
  {Schaefer}}]{Bluhm:2016byc}%
  \BibitemOpen
  \bibfield  {author} {\bibinfo {author} {\bibfnamefont {M.}~\bibnamefont
  {Bluhm}}, \bibinfo {author} {\bibfnamefont {M.}~\bibnamefont {Nahrgang}},
  \bibinfo {author} {\bibfnamefont {S.~A.}\ \bibnamefont {Bass}}, \ and\
  \bibinfo {author} {\bibfnamefont {T.}~\bibnamefont {Schaefer}},\ }\bibfield
  {title} {\enquote {\bibinfo {title} {{Impact of resonance decays on critical
  point signals in net-proton fluctuations}},}\ }\href
  {http://dx.doi.org/10.1140/epjc/s10052-017-4771-3} {\bibfield  {journal}
  {\bibinfo  {journal} {Eur. Phys. J. C}\ }\textbf {\bibinfo {volume} {77}},\
  \bibinfo {pages} {210} (\bibinfo {year} {2017})},\ \Eprint
  {http://arxiv.org/abs/1612.03889} {arXiv:1612.03889 [nucl-th]} \BibitemShut
  {NoStop}%
%%CITATION = ARXIV:1612.03889;%%
\bibitem [{\citenamefont {Mishra}\ \emph {et~al.}(2016)\citenamefont {Mishra},
  \citenamefont {Garg}, \citenamefont {Netrakanti},\ and\ \citenamefont
  {Mohanty}}]{Mishra2016}%
  \BibitemOpen
  \bibfield  {author} {\bibinfo {author} {\bibfnamefont {D.~K.}\ \bibnamefont
  {Mishra}}, \bibinfo {author} {\bibfnamefont {P.}~\bibnamefont {Garg}},
  \bibinfo {author} {\bibfnamefont {P.~K.}\ \bibnamefont {Netrakanti}}, \ and\
  \bibinfo {author} {\bibfnamefont {A.~K.}\ \bibnamefont {Mohanty}},\
  }\bibfield  {title} {\enquote {\bibinfo {title} {{Effect of resonance decay
  on conserved number fluctuations in a hadron resonance gas model}},}\ }\href
  {http://dx.doi.org/10.1103/PhysRevC.94.014905} {\bibfield  {journal}
  {\bibinfo  {journal} {Phys. Rev. C}\ }\textbf {\bibinfo {volume} {94}},\
  \bibinfo {pages} {014905} (\bibinfo {year} {2016})},\ \Eprint
  {http://arxiv.org/abs/1607.01875} {arXiv:1607.01875 [hep-ph]} \BibitemShut
  {NoStop}%
%%CITATION = ARXIV:1607.01875;%%
\bibitem [{\citenamefont {Nahrgang}\ \emph {et~al.}(2015)\citenamefont
  {Nahrgang}, \citenamefont {Bluhm}, \citenamefont {Alba}, \citenamefont
  {Bellwied},\ and\ \citenamefont {Ratti}}]{Nahrgang2015}%
  \BibitemOpen
  \bibfield  {author} {\bibinfo {author} {\bibfnamefont {M.}~\bibnamefont
  {Nahrgang}}, \bibinfo {author} {\bibfnamefont {M.}~\bibnamefont {Bluhm}},
  \bibinfo {author} {\bibfnamefont {P.}~\bibnamefont {Alba}}, \bibinfo {author}
  {\bibfnamefont {R.}~\bibnamefont {Bellwied}}, \ and\ \bibinfo {author}
  {\bibfnamefont {C.}~\bibnamefont {Ratti}},\ }\bibfield  {title} {\enquote
  {\bibinfo {title} {{Impact of resonance regeneration and decay on the
  net-proton fluctuations in a hadron resonance gas}},}\ }\href
  {http://dx.doi.org/10.1140/epjc/s10052-015-3775-0} {\bibfield  {journal}
  {\bibinfo  {journal} {Eur. Phys. J. C}\ }\textbf {\bibinfo {volume} {75}},\
  \bibinfo {pages} {573} (\bibinfo {year} {2015})},\ \Eprint
  {http://arxiv.org/abs/1402.1238} {arXiv:1402.1238 [hep-ph]} \BibitemShut
  {NoStop}%
%%CITATION = ARXIV:1402.1238;%%
\bibitem [{\citenamefont {Begun}\ \emph {et~al.}(2006)\citenamefont {Begun},
  \citenamefont {Gorenstein}, \citenamefont {Hauer}, \citenamefont
  {Konchakovski},\ and\ \citenamefont {Zozulya}}]{Begun2006}%
  \BibitemOpen
  \bibfield  {author} {\bibinfo {author} {\bibfnamefont {V.~V.}\ \bibnamefont
  {Begun}}, \bibinfo {author} {\bibfnamefont {M.~I.}\ \bibnamefont
  {Gorenstein}}, \bibinfo {author} {\bibfnamefont {M.}~\bibnamefont {Hauer}},
  \bibinfo {author} {\bibfnamefont {V.~P.}\ \bibnamefont {Konchakovski}}, \
  and\ \bibinfo {author} {\bibfnamefont {O.~S.}\ \bibnamefont {Zozulya}},\
  }\bibfield  {title} {\enquote {\bibinfo {title} {{Multiplicity Fluctuations
  in Hadron-Resonance Gas}},}\ }\href
  {http://dx.doi.org/10.1103/PhysRevC.74.044903} {\bibfield  {journal}
  {\bibinfo  {journal} {Phys. Rev. C}\ }\textbf {\bibinfo {volume} {74}},\
  \bibinfo {pages} {044903} (\bibinfo {year} {2006})},\ \Eprint
  {http://arxiv.org/abs/nucl-th/0606036} {arXiv:nucl-th/0606036 [nucl-th]}
  \BibitemShut {NoStop}%
%%CITATION = NUCL-TH/0606036;%%
\bibitem [{\citenamefont {Jeon}\ and\ \citenamefont {Koch}(1999)}]{Jeon1999}%
  \BibitemOpen
  \bibfield  {author} {\bibinfo {author} {\bibfnamefont {S.}~\bibnamefont
  {Jeon}}\ and\ \bibinfo {author} {\bibfnamefont {V.}~\bibnamefont {Koch}},\
  }\bibfield  {title} {\enquote {\bibinfo {title} {{Fluctuations of particle
  ratios and the abundance of hadronic resonances}},}\ }\href
  {http://dx.doi.org/10.1103/PhysRevLett.83.5435} {\bibfield  {journal}
  {\bibinfo  {journal} {Phys. Rev. Lett.}\ }\textbf {\bibinfo {volume} {83}},\
  \bibinfo {pages} {5435--5438} (\bibinfo {year} {1999})},\ \Eprint
  {http://arxiv.org/abs/nucl-th/9906074} {arXiv:nucl-th/9906074 [nucl-th]}
  \BibitemShut {NoStop}%
%%CITATION = NUCL-TH/9906074;%%
\bibitem [{\citenamefont {Bzdak}\ \emph {et~al.}(2013)\citenamefont {Bzdak},
  \citenamefont {Koch},\ and\ \citenamefont {Skokov}}]{Bzdak2013}%
  \BibitemOpen
  \bibfield  {author} {\bibinfo {author} {\bibfnamefont {A.}~\bibnamefont
  {Bzdak}}, \bibinfo {author} {\bibfnamefont {V.}~\bibnamefont {Koch}}, \ and\
  \bibinfo {author} {\bibfnamefont {V.}~\bibnamefont {Skokov}},\ }\bibfield
  {title} {\enquote {\bibinfo {title} {{Baryon number conservation and the
  cumulants of the net proton distribution}},}\ }\href
  {http://dx.doi.org/10.1103/PhysRevC.87.014901} {\bibfield  {journal}
  {\bibinfo  {journal} {Phys. Rev.}\ }\textbf {\bibinfo {volume} {C87}},\
  \bibinfo {pages} {014901} (\bibinfo {year} {2013})},\ \Eprint
  {http://arxiv.org/abs/1203.4529} {arXiv:1203.4529 [hep-ph]} \BibitemShut
  {NoStop}%
%%CITATION = ARXIV:1203.4529;%%
\bibitem [{\citenamefont {Begun}\ \emph {et~al.}(2004)\citenamefont {Begun},
  \citenamefont {Gazdzicki}, \citenamefont {Gorenstein},\ and\ \citenamefont
  {Zozulya}}]{Begun2004}%
  \BibitemOpen
  \bibfield  {author} {\bibinfo {author} {\bibfnamefont {V.~V.}\ \bibnamefont
  {Begun}}, \bibinfo {author} {\bibfnamefont {M.}~\bibnamefont {Gazdzicki}},
  \bibinfo {author} {\bibfnamefont {M.~I.}\ \bibnamefont {Gorenstein}}, \ and\
  \bibinfo {author} {\bibfnamefont {O.~S.}\ \bibnamefont {Zozulya}},\
  }\bibfield  {title} {\enquote {\bibinfo {title} {{Particle number
  fluctuations in canonical ensemble}},}\ }\href
  {http://dx.doi.org/10.1103/PhysRevC.70.034901} {\bibfield  {journal}
  {\bibinfo  {journal} {Phys. Rev. C}\ }\textbf {\bibinfo {volume} {70}},\
  \bibinfo {pages} {034901} (\bibinfo {year} {2004})},\ \Eprint
  {http://arxiv.org/abs/nucl-th/0404056} {arXiv:nucl-th/0404056 [nucl-th]}
  \BibitemShut {NoStop}%
%%CITATION = NUCL-TH/0404056;%%
\end{thebibliography}%

\end{document}